\documentclass[11pt]{article}

\usepackage{graphicx}
\usepackage{calc}
\usepackage{rotating}
\usepackage[english]{babel}
\usepackage{graphicx}
\usepackage{subfig}
\usepackage{float}
\usepackage{amsmath}
\usepackage{amssymb}
\usepackage{amsthm}
\usepackage{latexsym}
\usepackage{dcolumn}
\usepackage{geometry}
\usepackage{color}
\usepackage{hyperref}
\usepackage{mciteplus}
\usepackage{slashed}
\usepackage{multirow}

\def\gtwid{\mathrel{\raise.3ex\hbox{$>$\kern-.75em\lower1ex\hbox{$\sim
$}}}}
\def\vio{\mathrel{\hbox{$E$\kern-.60em\hbox{$/
$}}}}
\newcommand{\hobs}{\ensuremath{h_{\rm obs}}}
\newcommand{\omegan}{\ensuremath{\Omega_{\widetilde{\chi}^0_{1}}}}     
\newcommand{\omegadm}{\ensuremath{\Omega_{{\rm DM}}}}     
\newcommand{\neut}[1]{\ensuremath{\widetilde{\chi}^0_{#1}}}

\begin{document}

\begin{flushright}
KIAS-P18038\\
\end{flushright}
\vspace*{2.0cm}
\begin{center}
{\Large \bf {Quantum interference among heavy NMSSM Higgs bosons} \\
\vspace*{0.8cm}
{\large Biswaranjan Das$^a$, Stefano Moretti$^b$, 
Shoaib Munir$^c$ and Poulose Poulose$^a$ } \\[0.25cm]
{\small \sl $^a$Department of Physics, IIT Guwahati, Assam 781039, India} \\[0.25cm]
{\small \sl $^b$School of Physics \& Astronomy,
University of Southampton, \\ Highfield, Southampton 
SO17 1BJ, UK} \\[0.25cm]
{\small \sl $^c$School of Physics, Korea Institute for Advanced Study,
Seoul 130-722, Republic of Korea}\bigskip \\
{\small \url{biswaranjan@iitg.ernet.in}, \url{s.moretti@soton.ac.uk},
  \\ \url{smunir@kias.re.kr}, \url{poulose@iitg.ernet.in}}}
\end{center}
\vspace*{0.4cm}

\begin{abstract}
\noindent
In the Next-to-Minimal Supersymmetric 
Standard Model (NMSSM), it is possible to have strong 
mass degeneracies between the new singlet-like scalar and the heavy
doublet-like scalar, as well as between the singlet-like and doublet-like pseudoscalar Higgs states. When the difference in the masses of such states
is comparable with the sum of their widths, the quantum mechanical 
interference between their propagators can become significant.
We study these effects by taking into account the full Higgs boson propagator 
matrix in the calculation of the production process of $\tau^+\tau^-$ pairs
 in gluon fusion at the Large Hadron Collider (LHC).   
We find that, while these interference effects are sizeable, they are not 
resolvable in terms of the distributions of differential cross sections, owing to 
the poor detector resolution of the $\tau^+\tau^-$ invariant mass. They are, however, 
identifiable via the inclusive cross sections, which are subject to significant
variations with respect to the standard approaches, wherein the propagating Higgs bosons are treated independently from one another. We quantify these effects for 
several representative benchmark points, extracted from a large set of points, obtained by numerical scanning of the NMSSM parameter space, that satisfy the most important experimental constraints currently available.

\end{abstract}

\newpage
\section{Introduction}
\label{sec:intro}

Supersymmetry (SUSY) predicts the existence of at least 
two Higgs doublets of opposite hypercharge, which are 
necessary to cancel chiral anomalies, unlike the Standard 
Model (SM), where only one Higgs doublet is sufficient. In 
the Minimal Supersymmetric Standard Model (MSSM), after the 
breaking of the Electro-Weak (EW) symmetry, these two doublets 
result in a total of three physical neutral Higgs states, 
two scalars ($h$ and $H$, with $m_h < m_H$) and a 
pseudoscalar ($A$), as well as a charged pair ($H^\pm$). 
In principle, either of the $h$ and $H$ could be the Higgs 
boson, \hobs, observed at the LHC~\cite{Aad:2012tfa,Chatrchyan:2012xdj}. 
However, the model's parameter space region where $H$ has 
a mass near 125\,GeV and almost SM-like couplings to EW gauge 
bosons is very tightly constrained by  experimental 
data~\cite{Bechtle:2012jw,*Bechtle:2016kui}. Alternatively, the condition on $h$ to be 
a candidate for \hobs, while a phenomenologically much more favoured
scenario, pushes $m_H$ and $m_A$ upwards into 
the so-called decoupling regime (see, e.g.,~\cite{Djouadi:2005gj}), where they are 
nearly identical.

Thus, if a bump appears in the LHC data near a 
certain (large) invariant mass of a fermion-antifermion 
or photon pair above the SM expectation, it could very well 
be due to these two states combined, unless 
their mass difference is large enough to enable the experiment 
to resolve them individually. A formidable difference in the statistical 
significances of the peaks observed in the fermionic 
channels versus the $W^+W^-$ and $ZZ$ channels could 
also be interpreted as a hint of the $\{H,\,A\}$ mass-degeneracy. 
However, if the Higgs sector of the MSSM is CP-violating, 
the two heavy mass-degenerate states could mutually interfere 
quantum mechanically. This interference could significantly alter 
the cross section expected for a given SM final state that these Higgs
bosons decay into~\cite{Ellis:2004fs,Fuchs:2014ola,Fuchs:2016swt}.

In the NMSSM~\cite{Fayet:1974pd,*Ellis:1988er,*Durand:1988rg,*Drees:1988fc}, 
which originally aimed at addressing the $\mu$-problem of 
the MSSM~\cite{Kim:1983dt,Nir:1995bu} by invoking an additional $SU(2)_L$ singlet Higgs field, there exists an extra scalar, $h_s$, and an extra 
pseudoscalar, $a_s$, in the Higgs sector. The masses of 
these two new Higgs bosons are essentially free parameters 
of the model and either of them could very well lie near 125\,GeV,
along with the MSSM-like $h$.
The scenario with $m_{h_s}$ near $m_h$ is in fact well-motivated by
naturalness considerations, since the doublet-singlet 
mixing can enhance the tree-level mass of $h$ 
appreciably~\cite{Miller:2003ay,Ellwanger:2009dp,Maniatis:2009re}.
The potentially strong effects of 
interference in the scenario where these Higgs bosons are too close in mass
for the peaks to be separately 
identifiable, given the current diphoton mass resolution 
at the LHC, have therefore been recently analysed 
in~\cite{Das:2017tob}. In this work, we consider the alternative 
scenarios wherein one (or possibly even both) of $h_s$ and 
$a_s$ could be mass-degenerate with the heavier 
$H$ and $A$ instead of $h$. 
Our main objective here is to investigate the phenomenological implications
of such a scenario, which is a viable one in general extended Higgs sectors, 
without delving into its theoretical motivations in the specific model considered.

For our analysis, we first perform numerical scans of 
the NMSSM parameter space to identify regions where 
$m_{h_s}\approx m_H ~(\approx m_A)$ or 
$m_{a_s}\approx m_A~(\approx m_H)$. We then 
quantify the impact of quantum interference on the 
process where nearly mass-degenerate heavy Higgs bosons 
are produced in gluon fusion at the 14\,TeV LHC and decay 
into $\tau^+\tau^-$ pairs. This is done by comparing the 
cross section obtained by including the full Higgs 
propagator matrix in the expression for amplitude and the 
one obtained by assuming two (or more) individual 
Breit-Wigner (BW) Higgs propagators with nearly identical 
poles. We restrict ourselves to a CP-conserving NMSSM 
Higgs sector here, so that the scalar and pseudoscalar 
interaction eigenstates do not mix. Hence, the off-diagonal 
elements in the Higgs propagator matrix contribute to the 
interference affects only when the two mass-degenerate states 
have the same CP-identities. Furthermore, 
employing the diphoton final state, as in our previous study pertaining to 
125\,GeV Higgs bosons in a similar context~\cite{Das:2017tob}, is not feasible here. This is because the cross section, which is already phase-space suppressed, gets further depleted considerably due to the extremely small partial decay widths - and consequently Branching Ratios (BRs) - of the heavy Higgs states into photon pairs. 

During the numerical scanning, each randomly generated 
parameter space point is tested against the most crucial 
and recent experimental constraints, including those from 
the LHC, from $B$-physics measurements and from Dark Matter 
(DM) searches. Among the ones passing all the constraints, we 
then identify a few benchmark points (BPs) and carry out the cross section comparison 
noted above for them, using a Monte Carlo (MC) integration code 
developed in-house. We find that, as concluded by the previous 
studies for the MSSM~\cite{Fuchs:2016swt,Fuchs:2017wkq}, the interference is (almost) always 
destructive and can result in a sizeable reduction of the cross 
section for the considered process. However, the heavy Higgs 
boson masses for the successful points collected in our scans are always quite large ($>800$\,GeV). As a result, the effects of 
interference on the shape of the distribution of the differential cross
section may not be detectable even with an integrated luminosity of 
6000\,fb$^{-1}$ at the LHC, owing to the low $\tau^+\tau^-$ 
invariant mass resolution.

The article is organised as follows. In the next section we 
will briefly revisit the Higgs sector of the NMSSM and the 
analytical expression for the cross section that includes 
the full Higgs propagator matrix. In section \ref{sec:numerical} 
we will discuss in some detail our methodology for the 
parameter space scanning. We will present the 
results of our comparative numerical analysis of event rates in section \ref{sec:results} and our conclusions in section \ref{sec:concl}.

\section{The $\tau^+ \tau^-$ signal from heavy NMSSM Higgs bosons at the LHC}
\label{sec:cpvnmssm}

\subsection{\label{sec:higmat} The NMSSM Higgs sector}

The Higgs potential of the $Z_3$-symmetric NMSSM is written in terms of the two  $SU(2)_L$ doublets $H_u$ and $H_d$, with $Y = \pm 1$, and the singlet $S$ as
\begin{eqnarray}\label{eq:potential}
V_0 &=& {| \lambda \left(H_u^+ H_d^- - H_u^0 H_d^0 \right) + \kappa S^2 |}^2
+ \left(m_{H_u}^2 + {| \mu + \lambda S |}^2 \right)
  \left({| H_u^0 |}^2 + {| H_u^+ |}^2 \right) \nonumber \\ 
&+& \left(m_{H_d}^2 + {| \mu + \lambda S |}^2 \right)
  \left({| H_d^0 |}^2 + {| H_d^- |}^2 \right)
+ \frac{g^2}{4} \left({| H_u^0 |}^2 + {| H_u^+ |}^2
- {| H_d^0 |}^2 - {| H_d^- |}^2 \right)^2 \nonumber \\
&+& \frac{g_2^2}{2} {| H_u^+ H_d^{0*} + H_u^0 H_d^{-*} |}^2
+ m_S^2 {| S |}^2 + \left[ \lambda A_{\lambda} \left(H_u^+ H_d^- - H_u^0 H_d^0 \right) S
+ \frac{1}{3} \kappa A_{\kappa} S^3 + {\rm h.c.} \right].
\end{eqnarray}
Here $\lambda$ and $\kappa$ are dimensionless Higgs trilinear 
couplings and $A_{\lambda}$ and $A_{\kappa}$ are
their respective soft SUSY-breaking counterparts, 
$m_{H_d}$, $m_{H_u}$ and $m_S$ are the soft Higgs masses,
and $g_1$ and $g_2$ are the $U(1)_Y$ and $SU(2)_L$ gauge 
coupling constants, respectively, with $g^2 = \frac{g_1^2+g_2^2}{2}$.

By taking the second derivative of $V_0$ after developing the fields $H_d$, $H_u$ and $S$ around their respective Vacuum Expectation Values (VEVs), $v_d$, $v_u$ and $v_s$, as
\begin{equation}\label{eq:fields}
H_d^0 = 
\left( \begin{array}{c} \frac{1}{\sqrt 2}(v_d+H_{dR}+iH_{dI}) \\ H_d^- \end{array} \right),
H_u^0 = e^{i \phi_u}
\left( \begin{array}{c} H_u^+ \\ \frac{1}{\sqrt 2}(v_u+H_{uR}+iH_{uI}) \end{array} \right),
S^0 =  \frac{e^{i \phi_s}}{\sqrt 2}(v_s+S_R+iS_I)\,,
\end{equation}
one obtains the tree-level $5 \times 5$ neutral Higgs mass-squared matrix in the ${\rm H}^T = (H_{dR},H_{uR},S_R,H_I,S_I)$ basis, from which the Goldstone state, $G$, has been rotated away. In the CP-conserving limit, where all the Higgs sector coupling parameters are real, after the inclusion of the higher order corrections, this mass matrix can be expressed in the block-diagonal form 
\begin{equation}\label{eq:matrix}
{\cal M}_H^2 =
\left( \begin{array}{c|c} {\cal M}_S^2  &  \bf{0} \\  
\\  \hline & \\
  \bf{0} & {\cal M}_P^2 \end{array} \right)\,,
\end{equation}
where the $3 \times 3$ submatrix ${\cal M}_S^2$ corresponds to the CP-even states and the $2 \times 2$ matrix ${\cal M}_P^2$ to the CP-odd states, while all the CP-mixing terms vanish. The CP-even and CP-odd Higgs mass eigenstates can be obtained from the interaction states through the rotations
\begin{equation}\label{eq:mdiag}
\left( h_s,h,H \right)^T
= {\cal R}^H \left( H_{dR},H_{uR},S_R \right)^T ~~
{\rm and}~~\left( a_s,A \right)^T
= {\cal R}^A \left( H_I,S_I \right)^T,
\end{equation}
respectively, where ${\cal R}^H$ and ${\cal R}^A$ are orthogonal matrices.

Note that we have chosen the above notation for the physical Higgs states in order to distinguish between the MSSM-like and the NMSSM-specific ones. Throughout our analysis below, we identify the lighter MSSM-like (or doublet-like) scalar $h$ with the $\hobs$, so that the $H$ is always heavier than 125\,GeV, and we disregard the alternative possibility of $H$ playing the role of the \hobs. The mass of the singlet-like $h_s$ ($a_s$) can be smaller or larger than $m_h$ and/or $m_H$ ($m_A$). 

\subsection{\label{sec:XS} Gluon fusion production of Higgs bosons}

The differential cross section for the process $pp \to H \to \tau^+ \tau^-$ (with $H$ collectively denoting the five neutral Higgs bosons of the model, and assuming vanishing contribution from production modes other than gluon fusion) can be written as
\begin{equation}\label{eq:diffXShats}
\frac{d\sigma_{pp \to \tau^+\tau^-}}{d\sqrt{\hat s}} =
\int_{\tau}^{1} \frac{2\sqrt{\hat s}}{s} \frac{dx_1}{x_1}
\frac{g(x_1) g(\hat \tau x_1)}{1024 \pi \hat s} 
{\cal A}^2_{gg \to \tau^+\tau^-}\,,
\end{equation}
where $g(x_1)$ and $g(x_2)$ are the Parton Distribution Functions (PDFs) of the two incoming gluons having squared center-of-mass (CM) energy ${\hat s}$, and $\tau=\hat{s}/s$, with $s$ being the total CM energy of the $pp$ system. In the limit of negligible non-factorisable loop contributions, the amplitude-squared in the above equation can be cast into the most general form
\begin{equation}\label{eq:totamp}
{\cal A}^2_{gg \to \tau^+\tau^-} = \Big|\sum_{i,j=1-5} \sum_{\lambda,\sigma=\pm} 
{\cal M}_{P_i \lambda} D_{ij}{\cal M}_{D_j \sigma}\Big|^2\,,
\end{equation}
where $\lambda = \pm 1$ and $\sigma = \pm 1$ are the helicities of the incoming gluons and outgoing $\tau^\pm$, respectively. The matrix element for the production part, when the Higgs sector is CP-conserving, is given as 
\begin{equation}\label{eq:prodamp}
{\cal M}_{P_i \lambda} = \frac{\alpha_s \hat s}{4 \pi v} 
\Bigl\{ S_i^g + i \lambda P_i^g\Bigr\}\,, 
\end{equation}
where the scalar and pseudoscalar form factors, $S_i^g$ and $P_i^g$, respectively, can be found in, e.g.,~\cite{Lee:2003nta,Baglio:2013iia}. The amplitude for the decay part is similarly given as
\begin{equation}
  {\cal M}_{D_i \sigma}=\frac{\sqrt{2 {\hat s} \beta_{\tau}}\, m_{\tau}}{v} 
\Bigr\{\beta_{\tau} S_i^{\tau} + i \sigma P_i^{\tau} \Bigr\}\,,
\end{equation}
where $S_i^{\tau} = {\cal R}^H_{H_{dR} i}/\cos\beta$, $P_i^{\tau}=-{\cal R}^A_{H_I i}(\sin\beta/\cos\beta)$, and $\beta_{\tau}=\sqrt{1-4\frac{m_{\tau}^2} {\hat s}}$.

In principle, for calculating the differential cross section given in Eq.~\ref{eq:diffXShats}, one should include in the expression for ${\cal A}^2_{gg \to \tau^+\tau^-}$ the full NMSSM Higgs boson propagator matrix, written as 
\begin{equation}
\label{eq:propmat}
D_{ij} = 
\left(\begin{array}{@{}ccccc@{}} 
{\rm M_{11}} 
& i{{\mathfrak{I}}{\rm m}\hat\Pi}_{12}(\hat s) 
& i{{\mathfrak{I}}{\rm m}\hat\Pi}_{13}(\hat s)
& 0 
& 0 \\
i{{\mathfrak{I}}{\rm m}\hat\Pi}_{21}(\hat s) 
& {\rm M_{22}}
& i{{\mathfrak{I}}{\rm m}\hat\Pi}_{23}(\hat s)
& 0 
& 0 \\
i{{\mathfrak{I}}{\rm m}\hat\Pi}_{31}(\hat s) 
& i{{\mathfrak{I}}{\rm m}\hat\Pi}_{32}(\hat s) 
& {\rm M_{33}}
& 0 
& 0  \\
0
& 0 
& 0
& {\rm M_{44}}
& i{{\mathfrak{I}}{\rm m}\hat\Pi}_{45}(\hat s)  \\
0
& 0
& 0
& i{{\mathfrak{I}}{\rm m}\hat\Pi}_{54}(\hat s) 
& {\rm M_{55}}  \\
\end{array}\right)^{-1}\,, 
\end{equation}
where ${{\mathfrak{I}}{\rm m}\hat\Pi}_{ij}(\hat s)$ are the absorptive parts of the Higgs self-energies, for $i,\,j=1 - 5$, and ${\rm M}_{ii} \equiv \hat s-m_{H_i}^2 +i{{\mathfrak{I}}{\rm m}\hat\Pi}_{ii}(\hat s)$, with $m_{H_i}$ being the renormalised mass of the $i$th of the five neutral Higgs bosons. The explicit expressions for ${{\mathfrak{I}}{\rm m}\hat\Pi}_{ij}(\hat s)$ can be found in the Appendix of~\cite{Das:2017tob}. The total cross section can then be obtained by integrating Eq.~\ref{eq:diffXShats} over $\sqrt{\hat s}$. In general, however, the off-diagonal absorptive terms in the propagator are assumed to be negligible compared to all $m_{H_i}$. In that case, the amplitudes due to all $H_i$ can be summed {\it incoherently} as
\begin{equation}\label{eq:bwamp}
{\cal A}^2_{gg \to \tau^+\tau^-} = \sum_{i=1-5} \sum_{\lambda,\sigma=\pm} \Big|
{\cal M}_{P_i \lambda} \frac {1}{\hat s - m_{H_i}^2 +i{{\mathfrak{I}}{\rm m}\hat\Pi}_{ii}(\hat s)}  {\cal M}_{D_i \sigma}\Big|^2\,.
\end{equation}
In fact, if a given Higgs boson's width, $\Gamma_{H_i}$, defined through $i{{\mathfrak{I}}{\rm m}\hat\Pi}_{ii} \equiv im_{H_i}\Gamma_i$ from the corresponding diagonal term $M_{ii}$ in Eq.~\ref{eq:propmat}, is additionally much smaller than $m_{H_i}$, one can further apply the Narrow Width Approximation (NWA),
\begin{equation}\label{eq:Higgsprop}
{\left| \frac{1}{\hat s - m_{H_i}^2 + i m_{H_i} {\Gamma}_{H_i}} \right|^2} \to
\frac{\pi}{m_{H_i} {\Gamma}_{H_i}} \delta (\hat s - m_{H_i}^2)\,,
\end{equation}
to the amplitude above. This approximation implies 
on-shell production of each $H_i$, so that one can 
simply factorise its contribution to the total 
cross section as 
$\sigma_{pp\to H_i}\times {\rm BR}(H_i \to\tau^+\tau^-)$.

In the following, we define the total cross section 
for $\tau^+\tau^-$ production calculated using the above 
NWA as
\begin{equation}
  \label{eq:NWA}
  \sigma_{H_1...H_n }=\sum\limits_{H_i=H_1,...,H_n}\sigma (gg \to H_i)\times{\rm BR}(H_i\to\tau^+\tau^-)\,,
  \end{equation}
for all mass-degenerate $H_n$. This is the most commonly 
adopted approach, and we aim to examine to what extent 
this cross section might differ from the one obtained 
by the incoherent summing of amplitudes with individual BW 
propagators, as in Eq.~\ref{eq:bwamp}, which we refer 
to as $\sigma_{\rm BW}$. We then assess the further impact 
of the interference effects, by calculating the cross section, 
$\sigma_{\rm Int}$, calculated using the complete amplitude 
expression given in Eq.~\ref{eq:totamp}.  
 
\section{Parameter space: scans, constraints and favoured regions}
\label{sec:numerical}

In order to find solutions with nearly mass-degenerate $\{h_s,\,H\}$ and/or $\{a_s,\,A\}$ pairs that also satisfy experimental constraints from a variety of sources, we first performed numerical scanning of the NMSSM parameter space. As the model contains a large number of free parameters at the EW scale, we fixed the soft masses of the sfermions as $M_{Q_{1,2,3}} = M_{U_{1,2,3}} = M_{D_{1,2,3}} = 3$\,TeV and $M_{L_{1,2,3}} = M_{E_{1,2,3}}= 2$\,TeV, and of the gauginos as $2M_1 = M_2 = \frac{1}{3}M_3 = 1$\,TeV, since variations in these parameters are not expected to have a significant influence on the overall findings of our particular case study. Note also that the parameters $A_\lambda$ and $A_\kappa$ appearing in the potential in Eq.~\ref{eq:potential} can be traded for the pseudoscalar masses $m_P~(\sim m_{a_s})$ and $m_A$ as inputs, plus we assume universal trilinear couplings of the charged sfermions: $A_0 \equiv A_{\tilde{u},\tilde{c},\tilde{t}} = A_{\tilde{d},\tilde{s},\tilde{b}} = A_{\tilde{e},\tilde{\mu},\tilde{\tau}}$. The Higgs mass spectra and BRs for each randomly generated set of input parameters, over the wide initial scanned ranges of which are given in the second column of table~\ref{tab:scan}, were calculated using the public code {\tt NMSSMTools-v5.3.0}~\cite{NMSSMTools,Ellwanger:2004xm,*Ellwanger:2005dv}.

As for the experimental constraints, a scanned input point was rejected if it did not predict $m_h$ lying within the $123-127$\,GeV bracket (thus allowing for upto $\pm 2$\,GeV uncertainty in the theoretical prediction of its mass, given the experimental measurement of $125.09\pm 0.32$\,GeV~\cite{Aad:2015zhl}). In addition, we required the neutralino DM relic abundance, calculated by {\tt NMSSMTools} via an interface to {\tt MicrOmegas}~\cite{Belanger:2006is,Belanger:2014vza}, to satisfy $\omegan h^2 \leq 0.131$. This difference of the upper limit enforced on $\omegan h^2$ from the actual PLANCK measurement of $\omegadm = 0.119$~\cite{Planck:2015xua} is to accommodate up to a $+10\%$ possible error in its theoretical evaluation. A point was also discarded during the scan if the spin-independent \neut{1}-proton scattering cross section, $\sigma^{\rm SI}_p$, did not satisfy the 95\% Confidence Level (CL) limits from the XENON1T direct detection experiment~\cite{Aprile:2017iyp}. The theoretical estimate of this cross section is also written out in the {\tt NMSSMTools} output, as are those of the $B$-physics observables. The points collected in the scan were further filtered by the requirement on the most constraining of these observables to lie within 2$\sigma$ of their latest measurements, which read
\begin{itemize}
\item  ${\rm BR}(B\to X_s \gamma) \times 10^{4} = 3.32\pm0.15$~\cite{Amhis:2016xyh},
\item ${\rm BR}(B_u\to \tau^\pm \nu_\tau) \times 10^{4} = 1.06\pm0.19$~\cite{Amhis:2016xyh},
  \item ${\rm BR}(B_s \to \mu^+ \mu^-)\times 10^{9} = 3.0\pm 0.85$~\cite{Aaij:2017vad}.
\end{itemize}
    
\begin{table}[tbp]
\centering\begin{tabular}{|c|c|c|c|c|}
\hline
\multirow{2}{*}{Parameter} & Initial wide & Narrow range for & \multicolumn{2}{c|}{Narrow range for scenario 2 with} \\
 & scanned range & scenario 1 & $m_{h_s}<m_h$ &  $m_{h_s}>m_h$ \\
\hline
$A_0$\, (GeV)  & $-5000$ -- $-1000$ & $-5000$ -- $-3800$ & $-5000$ -- $-3800$ & $-5000$ -- $-1000$ \\
$\tan\beta$ 		& 2 -- 50 & 12 -- 17 & 2 -- 15 & 6 --17 \\
$\lambda$ 		& 0.001 -- 0.7 & 0.001 -- 0.02 &  0.01 -- 0.7  &  0.01 -- 0.3 \\
$\kappa$ 		& 0.001 -- 0.7 & 0.001 -- 0.04 & 0.01 -- 0.7 & 0.01 -- 0.7\\
$\mu_{\rm eff}$\,(GeV) & 100 -- 1000 & 145 -- 300 & 145 -- 250 & 145 -- 400\\
$m_A$\,(GeV)  	& 125 -- 1000 & 860 -- 1000 & 870 -- 1000 & 880 -- 1000 \\
$m_P$\,(GeV)  	& 10 -- 1000  & 10 -- 1000  & 880 -- 1000 & 890 -- 1000 \\
\hline
\end{tabular}
\caption{\label{tab:scan} Wide and narrowed-down (for a given
  scenario) scanned ranges of the seven NMSSM parameters considered
  free in this study.}
\end{table}

The successful points were then run through {\tt HiggsBounds-v4.3.1}~\cite{Bechtle:2008jh,*Bechtle:2011sb,*Bechtle:2013wla} to test each of the additional NMSSM Higgs bosons against the exclusion bounds from LEP, TeVatron and LHC as an added precaution, since this is done by {\tt NMSSMTools} itself also. These points were further subjected to the 95\% CL exclusion limits from the combined analysis of $\neut{1}\neut{1} \to b\bar{b}$ annihilation in dwarf spheroidal galaxies performed by the Fermi-LAT and MAGIC collaborations~\cite{Ahnen:2016qkx}, which are currently the strongest of the DM indirect detection bounds. Next, we calculated the theoretical predictions of the signal strengths, $\mu_X$, of $h$ in the SM decay channels, $X=\gamma\gamma,\,ZZ^*,\,W^+W^{-*},\,\tau^+\tau^-,\,b\bar{b}$, for each point using the public program {\tt HiggsSignals-v1.4.0}~\cite{Bechtle:2013xfa}. Since the discovery of \hobs, the CMS and ATLAS collaborations have frequently updated the measurements of $\mu_X$ independently from each other, and have also released their combined results based on the $\sqrt{s}=7$ and 8\,TeV data for each channel in~\cite{Khachatryan:2016vau}. However, no such combined analysis for the $\sqrt{s}=13$\,TeV data has been published so far and, while the measurements by the two groups have generally been in agreement with each other, there are also non-negligible differences in at least one of the channels (see, e.g.,~\cite{CMS:2018lkl} and~\cite{Aaboud:2017vzb,*Aaboud:2018xdt}). Given that these results have increasingly favoured the SM predictions, instead of choosing one of the two results over the other or performing fits to both, which would be beyond the scope of this study, we simply enforced $\mu_X = 1\pm 0.34$ for each $X$ on all points, i.e., the theoretical signal rates were required to lie within 1$\sigma$ of the SM value. 

Finally, it has previously been established in the literature~\cite{Fuchs:2017wkq,Cacciapaglia:2009ic} that the interference effects between two Higgs states grow as the mass-splitting between them drops off compared to the sum of their widths. We therefore defined
\begin{equation}
\Lambda_{X_i} = \frac{\Gamma_{X_i}}{\Delta m_X},~~{\rm with}~~\Delta m_X=m_{X_2}-m_{X_1}\,,
\end{equation}
where $i=1$ implies the lighter and $i=2$ the heavier of the two nearly mass-degenerate scalars ($X=H$) or pseudoscalars ($X=A$), and retained only the points for which $\Lambda_{X_1}>1$ or $\Lambda_{X_2}>1$. These points were then split into two categories,
\begin{itemize}
\item scenario-1: $m_{h_s}\approx m_H$, 
\item scenario-2: $m_{a_s} \approx m_A$.
  \end{itemize}
Note, however, that for a vast majority of the points left after applying the above filters, the doublet-like $H$ and $A$ were found to be highly mass-degenerate and thus lying in the decoupling regime of the MSSM. This is understandable, in particular in scenario-1, since the requirement for $h_s$ to have a mass close to that of $H$ forces $h$ to be almost entirely doublet-like, with maximal tree-level mass and SM-like couplings. Furthermore, for all the points the lightest neutralino turned out to be a pure higgsino, with the difference in its mass from that of the next-to-lightest neutralino always lying within $10-13$\,GeV. A recent ATLAS search~\cite{Aaboud:2017leg} has ruled out such a DM for a mass up to $\sim 145$\,GeV (which effectively translates into $\mu_{\rm eff} \ge 145$\,GeV). We therefore removed all the points with $m_{\widetilde{\chi}^0_1} < 145$\,GeV from the scanned set. We point out here that the limits from a CMS search~\cite{Sirunyan:2018ubx} which could also be of relevance here have already been incorporated in {\tt NMSSMTools-v5.3.0}, so that each scanned point was intrinsically tested against them. 

The points in scenario-2 can be further divided into two distinct sets, one with $h_s$ lighter than $h$ and the other with $h_s$ heavier than $h$. Columns 3--5 of table~\ref{tab:scan} show the ranges of the input parameters covered by the points corresponding to the two scenarios. In order to find more solutions with a possibly enhanced mass-degeneracy in a given scenario, we performed secondary scans of each of these narrowed-down parameters ranges. All the constraints and conditions noted above were reapplied to the points collected in these scans and the successful ones were combined with the corresponding set from the initial scan. 

\begin{figure}[t!]
    \begin{tabular}{cc}
\includegraphics*[width=7.7cm]{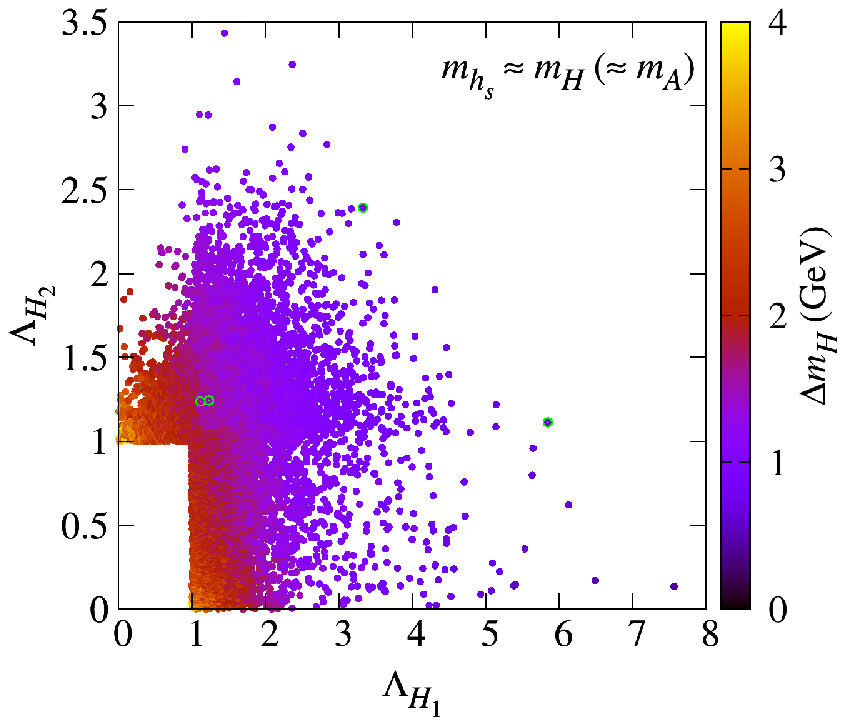} &
\includegraphics*[width=8.2cm]{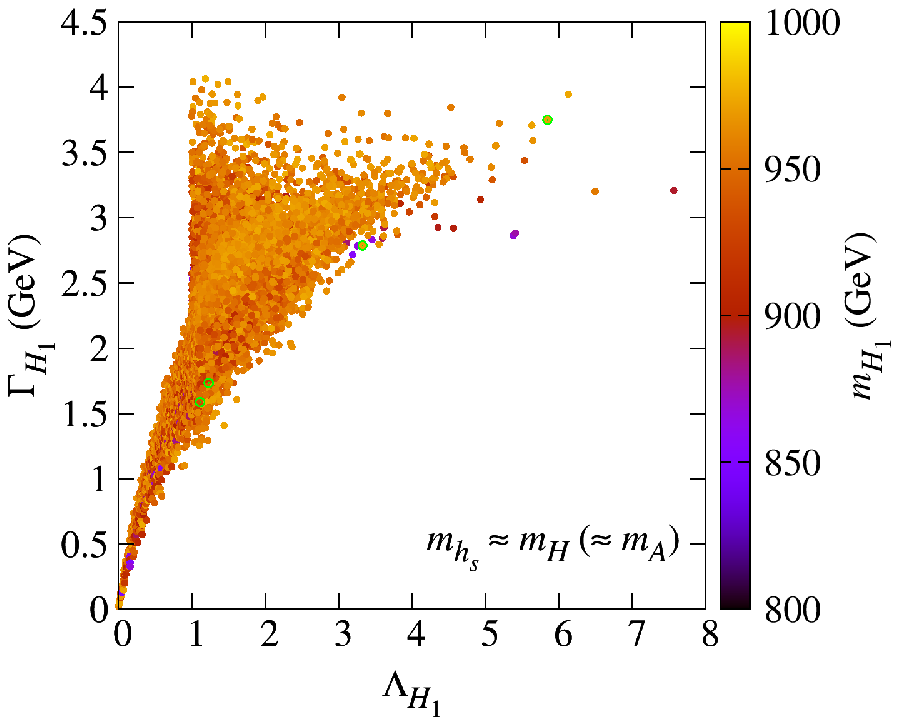} \\
\includegraphics*[width=8cm]{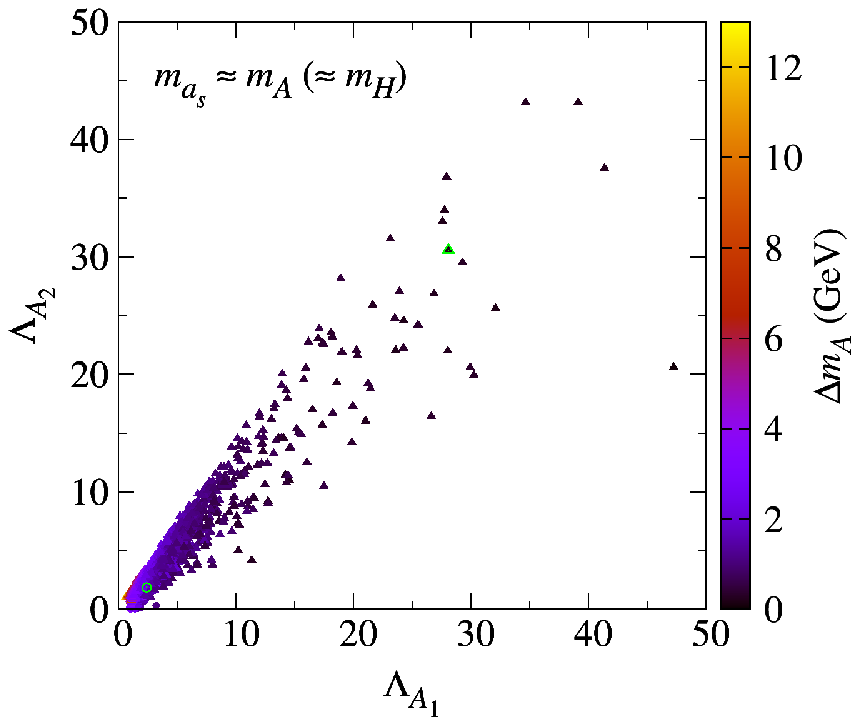} &
\includegraphics*[width=8.2cm]{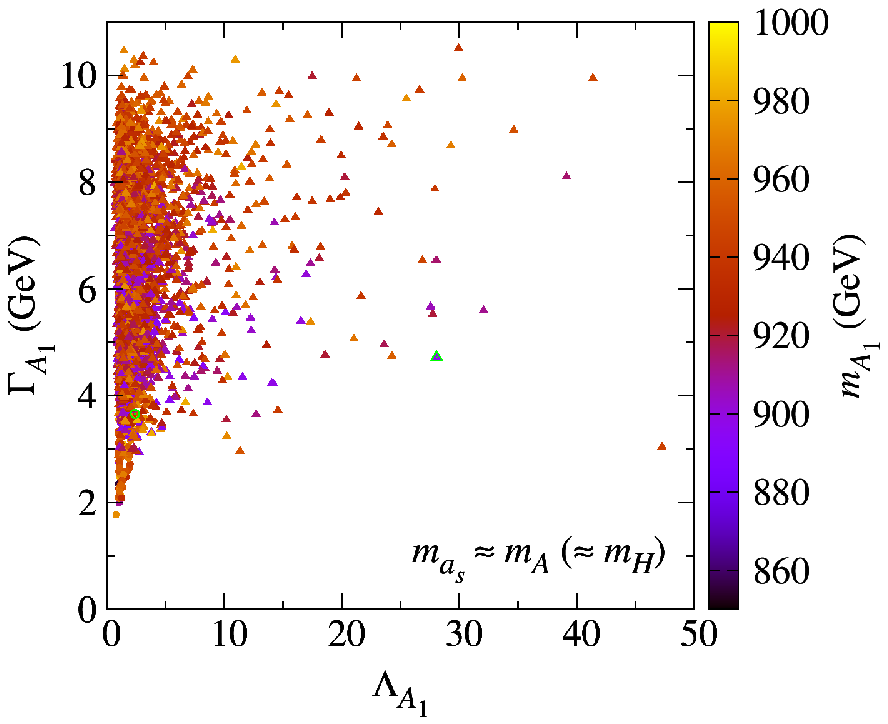}
\end{tabular}
\caption{\label{fig:lambda-width} $\Lambda_{X_i}$ vs. $\Delta m_X$
  (left), and $\Gamma_{X_i}$ vs. $m_{X_1}$ (right), for the points corresponding to
  scenario-1 (top) and scenario-2 (bottom), obtained from the parameter 
space scans of the NMSSM. See text for details.}
\end{figure}

In the top left panel of figure~\ref{fig:lambda-width}, we show $\Lambda_{H_i}$ for the final set of points belonging to scenario-1 (where $H_1$, and likewise $H_2$, can be either one of $h_s$ or $H$). We see that $\Gamma_{H_1}$ can be up to 8 times larger than $\Delta m_H$, as illustrated by the colour map. According to the top right panel, the maximum width of $H_1$ for these points is about 4\,GeV (while that of $H_2$, not shown here, is much smaller, as can be deduced from the overall smaller $\Lambda_{H_2}$ in the left panel). We point out again that $\Lambda_{H_2} > 1$ whenever $\Lambda_{H_1}<1$, and vice versa. In the bottom left panel of figure~\ref{fig:lambda-width}, which similarly shows $\Lambda_{A_i}$ for scenario-2, one notices $\Gamma_{A_1}$ being as much as 50 times larger than $\Delta m_A$. In this figure (and in the subsequent figures for this scenario) circles correspond to the subset of points for which $m_{h_s}>m_h$ and triangles to those with $m_{h_s}<m_h$. Hence, while in the former case $\Lambda_{A_2}$ never exceeds $\Lambda_{A_1}$, there are a few of the latter points for which both $\Lambda_{A_1},\,\Lambda_{A_1}\sim 40$. The bottom right panel illustrates that in the $m_{h_s}>m_h$ case large values of $\Lambda_{A_1}$ are a consequence of large, $\sim 10$\,GeV, values of the $A_1$ width. In the $m_{h_s}<m_h$ case, in contrast, $\Gamma_{A_1}$ stays low generally (between $2 - 4$\,GeV), and large $\Lambda_{A_1}$ results mainly from the fact that $\Delta m_A$ can reach values lower than in the $m_{h_s}>m_h$ case, according to the colour map in the bottom left panel. The points encircled in green in figure~\ref{fig:lambda-width} are the BPs we identified for our cross section analysis, to be explained in the next section. 

\begin{figure}[t!]
\begin{tabular}{cc}
\includegraphics*[width=8.2cm]{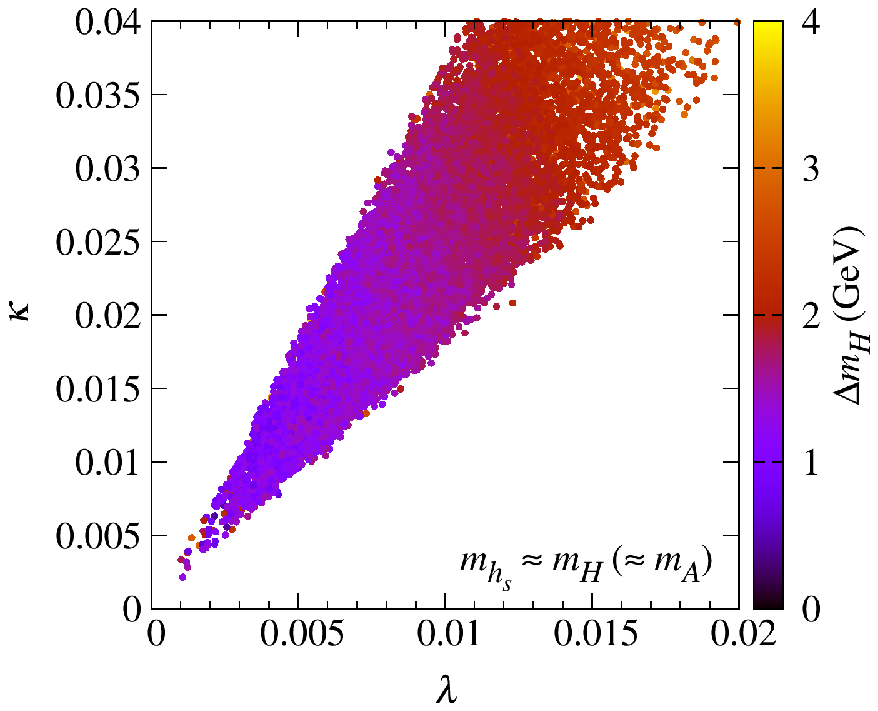} &
\includegraphics*[width=8cm]{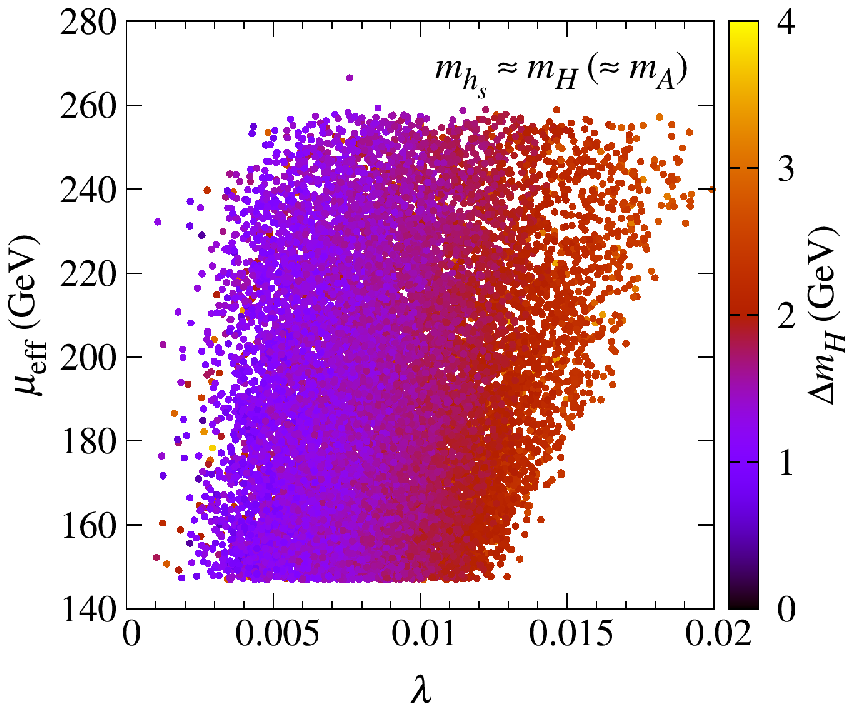} \\
\includegraphics*[width=8.2cm]{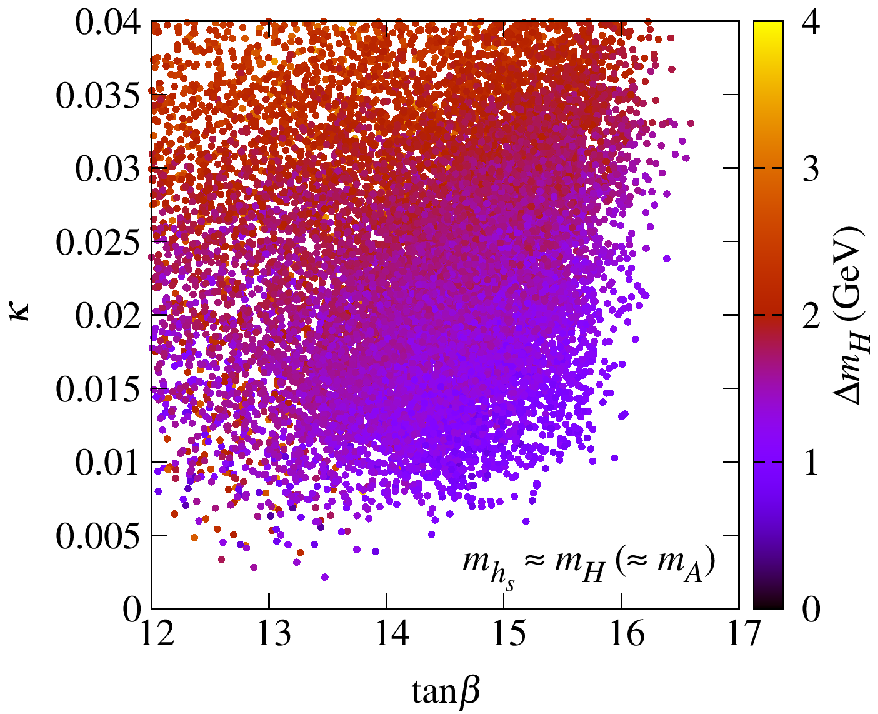} &
\includegraphics*[width=8cm]{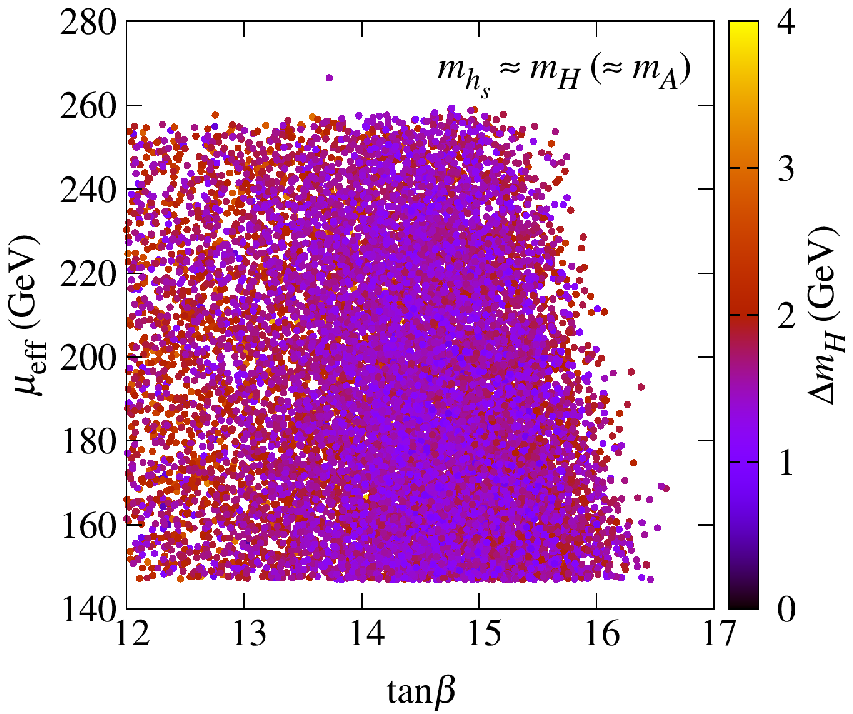}
\end{tabular}
\caption{\label{fig:Scen1} Distributions of the input parameters,
  showing the correlations among them that lead to a strong mass-degeneracy between $h_s$ and $H$, for the points corresponding to scenario-1 obtained from the numerical scans.}
\end{figure}

We now briefly discuss the parameter combinations that lead to strong mass-degeneracies between Higgs bosons in the two scenarios. Figure~\ref{fig:Scen1} corresponds to scenario-1, and shows that a smaller $\Delta m_H$ favours lower values of both $\lambda$ and $\kappa$ but larger values of $\tan\beta$ (within the considered range). The reason is that the mass-degeneracy condition in this scenario effectively implies the decoupling of $h_s$ from $h$ too. The smaller values of $\lambda$ essential for that in turn put a virtual cut-off on $\mu_{\rm eff} \equiv \lambda v_s$ of about 250\,GeV, below which it can vary relatively freely.

\begin{figure}[t!]
\begin{tabular}{cc}
  \includegraphics*[width=8cm]{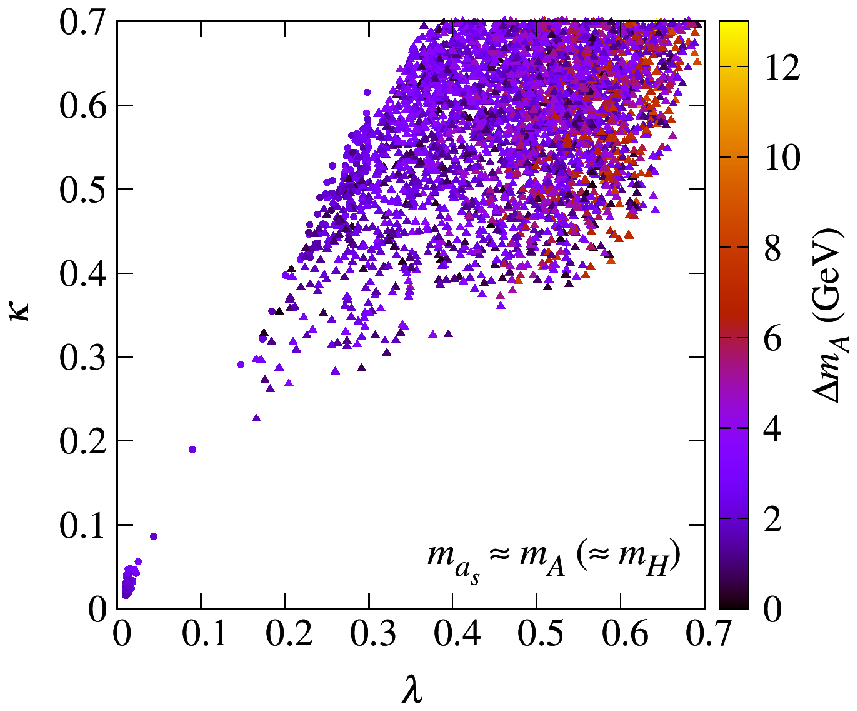} &
\includegraphics*[width=8cm]{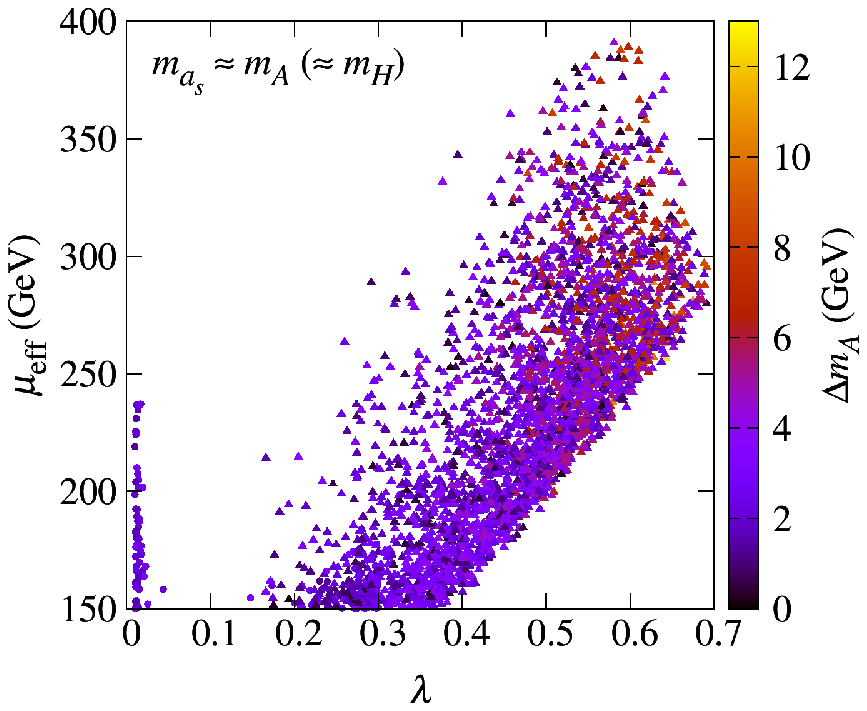} \\
\includegraphics*[width=8cm]{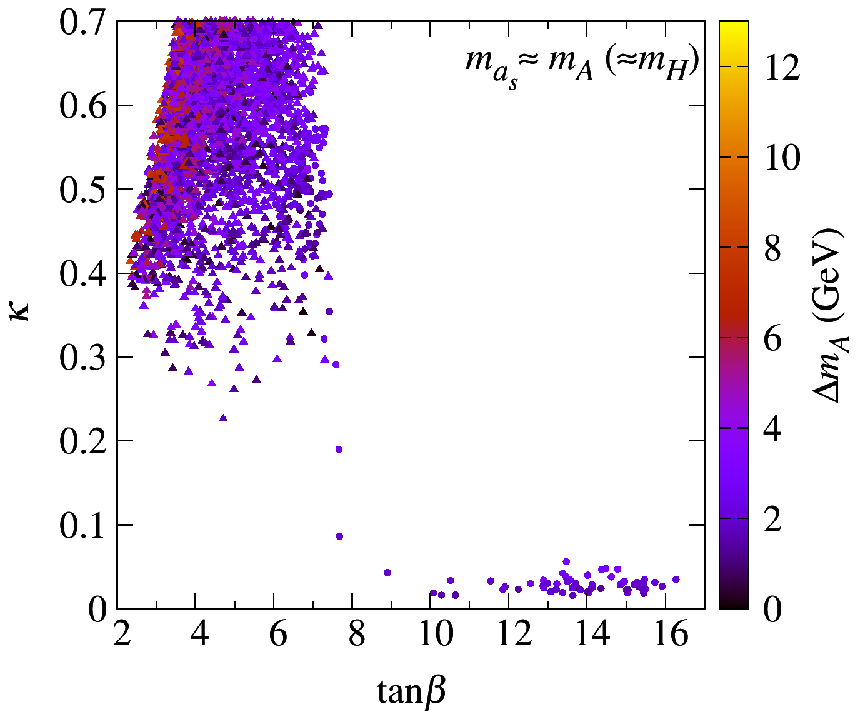} &
\includegraphics*[width=8cm]{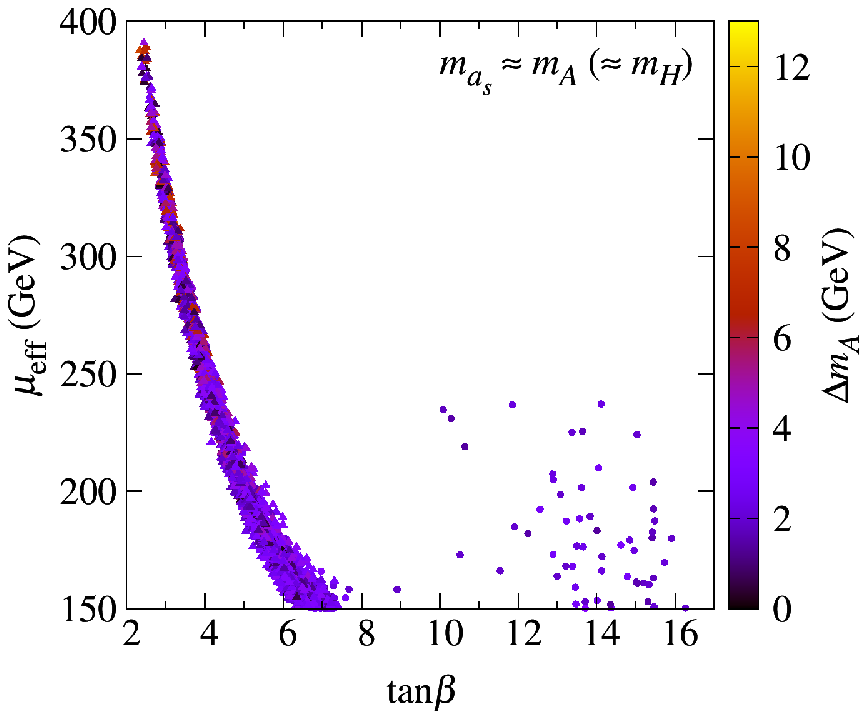}
\end{tabular}
\caption{\label{fig:Scen2} Distributions of the input parameters,
  showing the correlations among them that lead to a strong mass-degeneracy between $a_s$ and $A$, for the points corresponding to scenario-2 obtained from the numerical scans.}
\end{figure}

Scenario-2 shows a much stronger dependence on the correlations between the input parameters, as seen in figure~\ref{fig:Scen2}. In particular, the parameter space regions with smaller $\Delta m_A$ are rather distinct for the $m_{h_s}<m_h$ and $m_{h_s}>m_h$ cases. Recall that the $m_{h_s}<m_h$ case forces the model into the `natural' limit, so that both $h$ and $h_s$ are states with a large singlet-doublet mixing, made possible by larger $\lambda$ and $\kappa$, and smaller $\tan\beta$, as noted for the triangles in the various panels of the figure. Such values of these parameters, with some fine tuning, allow $\mu_{\rm eff}$ to reach up to $\sim 400$\,GeV. Note, however, that in the $m_{h_s}>m_h$ case also it is possible for $h_s$ to be a highly mixed state lying very close in mass to the $h$, and for such points the parameter values can be similar to the ones in the $m_{h_s}<m_h$ case. Generally though, the parameter combinations in the $m_{h_s}>m_h$ case follow a somewhat similar trend to that seen for scenario-1, with relatively small (large) $\lambda$ and $\kappa$ ($\tan\beta$), and $\mu_{\rm eff}$ restricted to below 250\,GeV. 

\section{Cross section analysis}
\label{sec:results}

As noted earlier, we selected some BPs from the two scenarios to study the impact of quantum interference between the Higgs bosons on the cross section for the $pp \to \tau^+\tau^-$ process at the LHC with $\sqrt{s}=14$\,TeV. The calculation of the cross sections $\sigma_{\rm BW}$ and $\sigma_{\rm Int}$ defined in section \ref{sec:XS} was carried out using a {\tt Fortran} program developed in-house, which performs Higgs propagator matrix inversion and numerical integration through interfaces with the {\tt LAPACK} package~\cite{lapack} and a locally modified version of the {\tt VEGAS} routine~\cite{Lepage:1977sw}, respectively. The Higgs boson masses and couplings for a given NMSSM parameter space point, which the program takes as inputs for the calculation of the propagator matrix as well as the form factors for Higgs boson production via gluon fusion, were written out by a suitably modified version of {\tt NMSSMTools}.

\begin{table}[t!]
\centering\begin{tabular}{|c|c|c|c|c|c|c|}
\hline
BP & 1 & 2 & 3 & 4 & 5 & 6 \\
\hline
\hline
$A_0$ (GeV) & -4624.6 & -4516.5 & -4371.9 & -4574.8 & -4967.9 &
-4518.8 \\
$\tan\beta$ & 13.90 & 13.84 & 15.14 & 15.84 & 6.42 & 5.65 \\
$\lambda$ & 0.0045 & 0.0034 & 0.0035 & 0.0041 & 0.2965 & 0.3948 \\
$\kappa$ & 0.0092 & 0.0068 & 0.0112 & 0.0141 & 0.5486 & 0.6197 \\
$\mu_{\rm eff}$ (GeV) & 217.34 & 217.73 & 150.50 & 152.63 &
151.21 & 172.92 \\
$m_A$ (GeV) & 926.92 & 904.00 & 994.13 & 998.86 & 898.56 &
902.80 \\
$m_P$ (GeV) & 72.37 & 698.12 & 189.83 & 626.85 & 919.23 &
931.95 \\
\hline
\hline
$m_{h}$ (GeV) & 124.13 & 124.16 & 123.84 & 124.23 &
123.04 & 124.21 \\
\hline
$m_{h_s}$ (GeV) & 889.98 & 893.37 & 970.47 & 973.01 &
191.07 & 107.13 \\
$m_{H}$ (GeV) & 891.39& 894.86 & 971.11 & 973.85 &
895.73 & 900.11 \\
$m_{a_s}$ (GeV) & 72.36 & 218.19 & 189.83 & 626.84 &
893.97 & 896.63 \\
$m_{A}$ (GeV) & 891.21 & 894.63 & 970.87 & 973.61 &
892.45 & 896.46 \\
\hline
\hline
$\Delta m_H$ (GeV) & 1.41 & 1.49 & 0.64 & 0.84 & & \\
$\Delta m_A$ (GeV) & & & & & 1.53 & 0.17 \\
\hline
\hline
$\Gamma_{h}$ (MeV) & 4.11 & 4.11 & 4.04 & 4.08 & 4.09 & 2.90 \\
\hline
$\Gamma_{h_s}$ (GeV) & 1.75 & 1.93 & 0.71 & 2.01 & & \\
$\Gamma_{H}$ (GeV) & 1.73 & 1.92 & 3.75 & 2.79 & 3.65 & 4.84 \\
$\Gamma_{a_s}$ (GeV) & & &  & & 2.86 & 5.14 \\
$\Gamma_{A}$ (GeV) & 3.53 & 3.87 & 4.49 & 4.82 & 3.65 &
4.72 \\
\hline
\hline
BR($h \to \tau^+\tau^-$) & 0.069 & 0.069 & 0.069 & 0.068 & 0.071 & 0.061 \\
\hline
BR($h_s \to \tau^+\tau^-$) & 0.103 & 0.102 & 0.103 & 0.106 & 0.005 & 0.091 \\
BR($H \to \tau^+\tau^-$) & 0.102 & 0.100 & 0.100 & 0.105 & 0.021 & 0.012 \\
BR($a_s \to \tau^+\tau^-$) & 0.087 & 0.012 & 0.010 & 0.002 & $10^{-5}$ & $10^{-7}$ \\
BR($A \to \tau^+\tau^-$) & 0.101 & 0.101 & 0.103 & 0.105 & 0.021 & 0.013 \\
\hline
\hline
$\sigma_{h_sHA}$ (fb)  & 0.547 & 0.537 & 0.334 & 0.322 & & \\
$\sigma_{Ha_sA}$ (fb)  & & & & & 0.364 & 0.267 \\
\hline
\hline
$\sigma_{\rm BW}$ (fb) & 0.637 & 0.584 & 0.354 & 0.351 &
0.445 & 0.314 \\
$\Delta \sigma_{\rm BW}$ (\%) & 16 & 9 & 6 & 9 & 22 & 17 \\ \hline
$\sigma_{\rm Int}$ (fb) & 0.565 & 0.514 & 0.314 & 0.286 
& 0.445 & 0.314 \\
$\Delta \sigma_{\rm Int}$ (\%)  & $-11$  & $-12$  & $-11$ & $-19$ & 0 & 0 \\

\hline
\end{tabular}
\caption{\label{Tab:BPdetails} Masses and decay widths of the Higgs bosons,
  and the cross sections obtained using the three approaches discussed
  in the text, for the six selected benchmark points. BPs 1--4
  correspond to scenario-1 and BPs 5 and 6 to scenario-2. Blank space in front of a quantity implies that it is not relevant for the given BP.}
\end{table}

For comparison, we also calculated $\sigma_{H_1...H_n}$ by multiplying the $\sigma(gg\to H_i)$ obtained from the public code {\tt SusHi v1.6.0}~\cite{Harlander:2012pb,*Liebler:2015bka,*Harlander:2016hcx} with BR$(H_i \to \tau^+\tau^-)$ given by {\tt NMSSMTools}. We note here that we required {\tt SusHi} to evaluate these $\sigma(gg\to H_i)$ at the Next-to-Next-to-Leading Order (NNLO) in QCD, while our program calculates the production process only at the Leading Order (LO).\footnote{Including higher order (HO) corrections for the production process is a highly involved task, which would be beyond the scope of this work. Their inclusion is, however, tantamount to rescaling the cross section for a given $H_i$ by a `$k$ factor', which is defined as $k_{\rm HO}\equiv \sigma_{\rm HO} / \sigma_{\rm LO}$, with HO implying the perturbative order at which the cross section is to be evaluated.} For a more accurate comparison, we therefore estimated the $k_{\rm NNLO}$ factor by obtaining also the LO $\sigma(gg\to H_i)$ from {\tt SusHi} for each $H_i$ in a given BP and multiplied our LO amplitude-squared with it before summing over all $H_i$'s to obtain the total $\sigma_{\rm BW}$. In the case of $\sigma_{\rm Int}$ though, since the contribution due to every $H_i$ is not computed separately, we simply multiply the total amplitude-squared with an average of the $k_{\rm NNLO}$ factors for the (three) Higgs bosons near a given $\tau^+\tau^-$ invariant mass (or, equivalently,  $\sqrt{\hat s}$). For all the cross sections, the CT10~\cite{Lai:2010vv} PDF sets were used for gluons and the default values of the SM input parameters inside {\tt NMSSMTools} were retained, except for the $t$-quark mass, which was fixed to 172.5\,GeV.

The values of the input parameters, the masses and widths of the Higgs bosons, as well as the total cross sections calculated using each of the approaches explained above are given in table~\ref{Tab:BPdetails} for all the selected BPs. BP1--BP4 correspond to scenario-1, and the first one of these has been selected such that $m_{h_s}$ and $m_{H}$ lie relatively close to their lowest observed values, in order to minimise phase-space suppression of the cross section, while also requiring $\Lambda_{H_{1,2}}>1$. However, these masses are still only slightly smaller than 900\,GeV, and thus, while the $\sigma_{H_1...H_n}$ for this BP is the highest among all, it nevertheless stays below 1\,fb. Also noted down in the table is the quantity $\Delta \sigma_{\rm BW} \equiv \frac {\sigma_{\rm BW} - \sigma_{H1...H_n}}{\sigma_{H1...H_n}}\times 100$ and its value for BP1 indicates that the total cross section assuming BW propagators is enhanced by 16\% over the one obtained using the NWA. $\Delta \sigma_{\rm Int} \equiv \frac {\sigma_{\rm Int} - \sigma_{\rm BW}}{\sigma_{\rm BW}}\times 100$  in the last row of the table similarly quantifies the impact of including the full propagator matrix in the amplitude. We note that $\sigma_{\rm Int}$ is 11\% smaller than $\sigma_{\rm BW}$, implying that the interference between the $h_s$ and $H$ propagators is destructive, reducing the total cross section somewhat, although it is still larger than $\sigma_{h_sHA}$ (i.e, $\sigma_{H1...H_n}$, defined in Eq.~\ref{eq:NWA}, calculated assuming the NWA for $H_1,...,H_n = h_s,\,H,\,A$). 

For all the points with $m_{h_s}$ and $m_{H}$ similar to those for BP1, $\Lambda_{H_{1,2}}$ stays very close to 1. Larger values of these ratios are typically obtained for higher masses, and BP2 was selected such that it has the former slightly increased, while the latter are still below 900\,GeV. Even this slight increase in mass reduces each cross section by a few percent compared to the BP1 case, while enhancing the interference effects only marginally. The interference for this BP, and in fact also for the other two BPs for scenario-1, is again destructive. The selection criteria for BP3 were maximal $\Lambda_{H_{1}}$ only, while for BP4 maximal $\Lambda_{H_{1,2}}$, with their respective $m_{h_s}$ and $m_H$ being very similar. This effectively implies that for BP3 one has $\Gamma_{H_{2}} (= \Gamma_{h_s})$ considerably smaller than $\Gamma_{H_{1}} (= \Gamma_H)$, while for BP4 both these widths are much closer in size. This difference leads to a clear rise in the magnitude of interference for BP4, which leads to a $\Delta \sigma_{\rm Int}$ of 19\%, compared to the BP3, for which the increase in the cross section is 11\%, identically to BPs 1 and 2. $\Delta \sigma_{\rm BW}$ for BPs 3 and 4 are also considerably smaller than for BP1, despite a large hike in $m_{h_s}$ and $m_H$, as is the respective absolute value of each type of the cross section itself. Note that $\mu_{\rm eff}$ for both these high-mass BPs lies just above the effective exclusion limit from ATLAS discussed earlier, and could thus be within the reach of subsequent analyses at the LHC in the near future.  BR($H_i \to \tau^+\tau^-$), with $H_i=h_s,\, H,\,A$, for all the above BPs is $\sim 10\%$.  

BP5 belongs to the $m_{h_s}>m_h$ case and BP6 to the $m_{h_s}<m_h$ case of scenario-2. As with BP3 and BP4, for BP5 also one finds that $\Gamma_{a_s}$ is somewhat smaller than $\Gamma_A$, while for BP6 the two widths are much closer in magnitude. Furthermore, $\sigma_{\rm BW}$ shows an enhancement of about 22\% over $\sigma_{Ha_sA}$ for BP5, but $\sigma_{\rm Int}$ is exactly the same as $\sigma_{\rm BW}$, implying that interference is non-existent between the $a_s$ and $A$ propagators. This is also the case for BP6, despite $\Lambda_{a_s}$ and $\Lambda_{A}$ both being very large due to a very strong mass-degeneracy between $a_s$ and $A$. $\sigma_{h_sHA}$ for BP6 is sizeably smaller than for BP5, which is a consequence of the relatively small BR($A \to \tau^+\tau^-$), even though the respective pseudoscalar masses are only about 4\,GeV larger. Notice that for both of these BPs, even though $\Gamma_{a_s}$ are of order 1\,GeV, the $\tau^+\tau^-$ partial decay widths, and hence the BR($a_s \to \tau^+\tau^-$), are extremely small, resulting in the vanishing of the interference effects.   

\begin{figure}[t!]
\begin{tabular}{cc}
\includegraphics*[width=7.8cm]{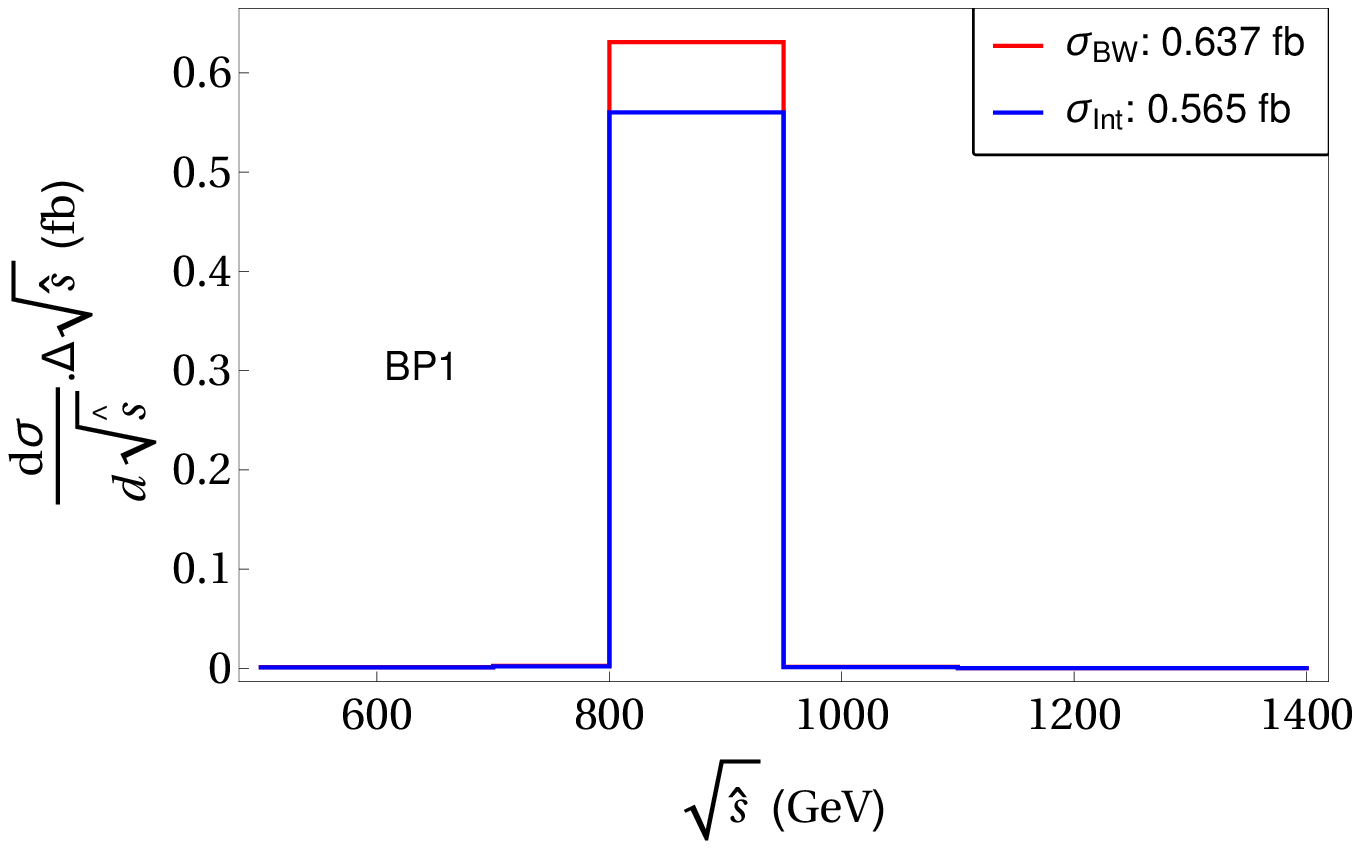} &
\includegraphics*[width=7.8cm]{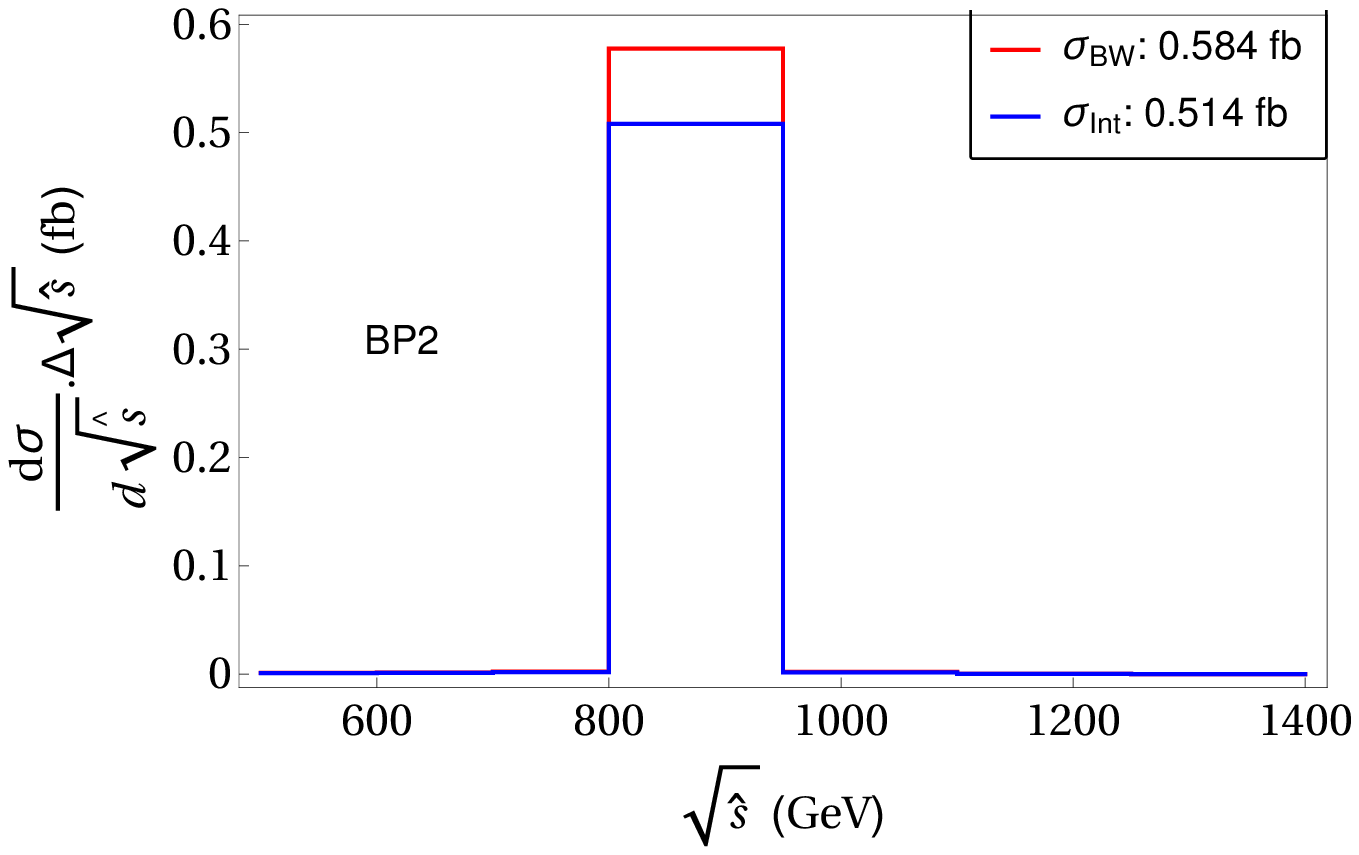} \\
\includegraphics*[width=7.8cm]{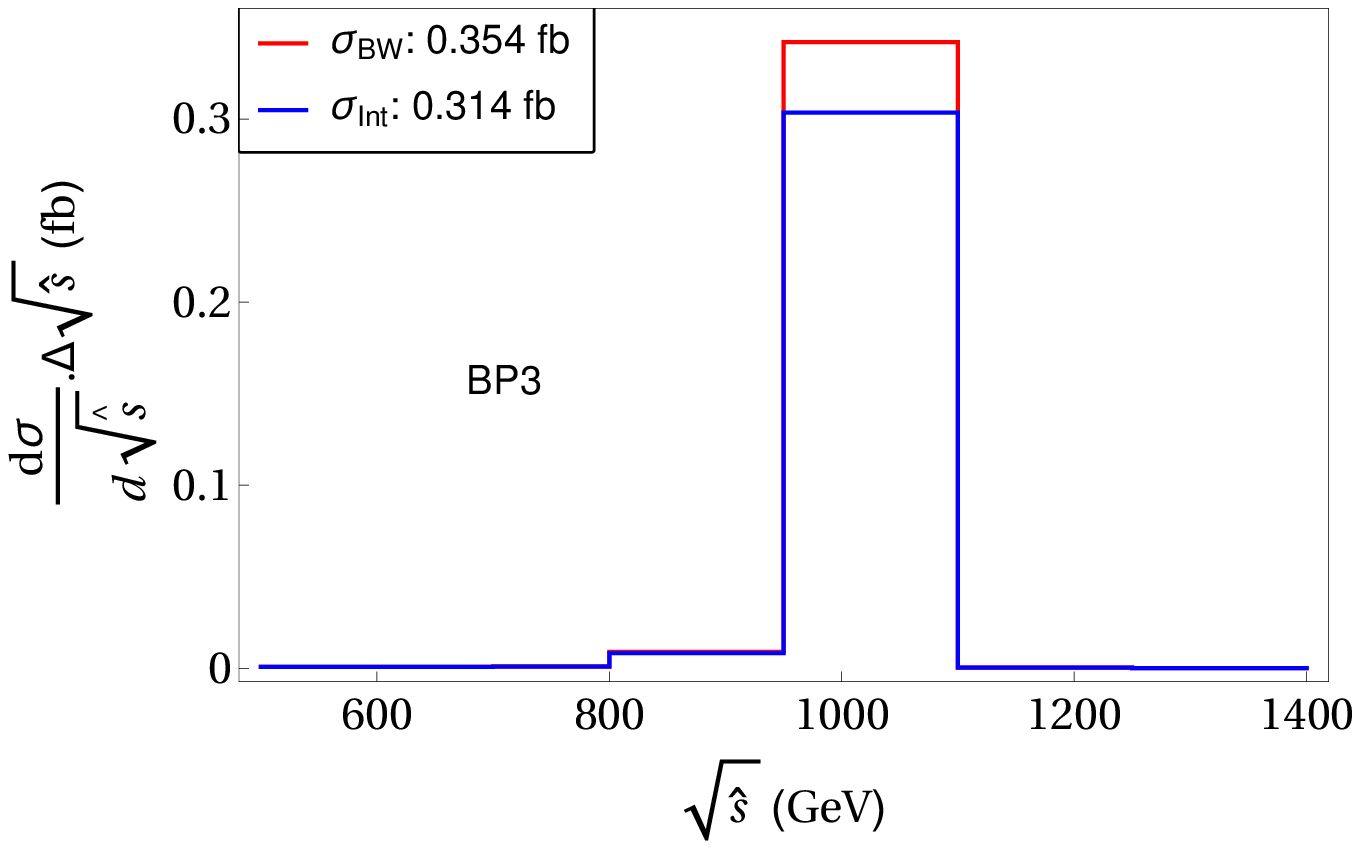} &
\includegraphics*[width=7.8cm]{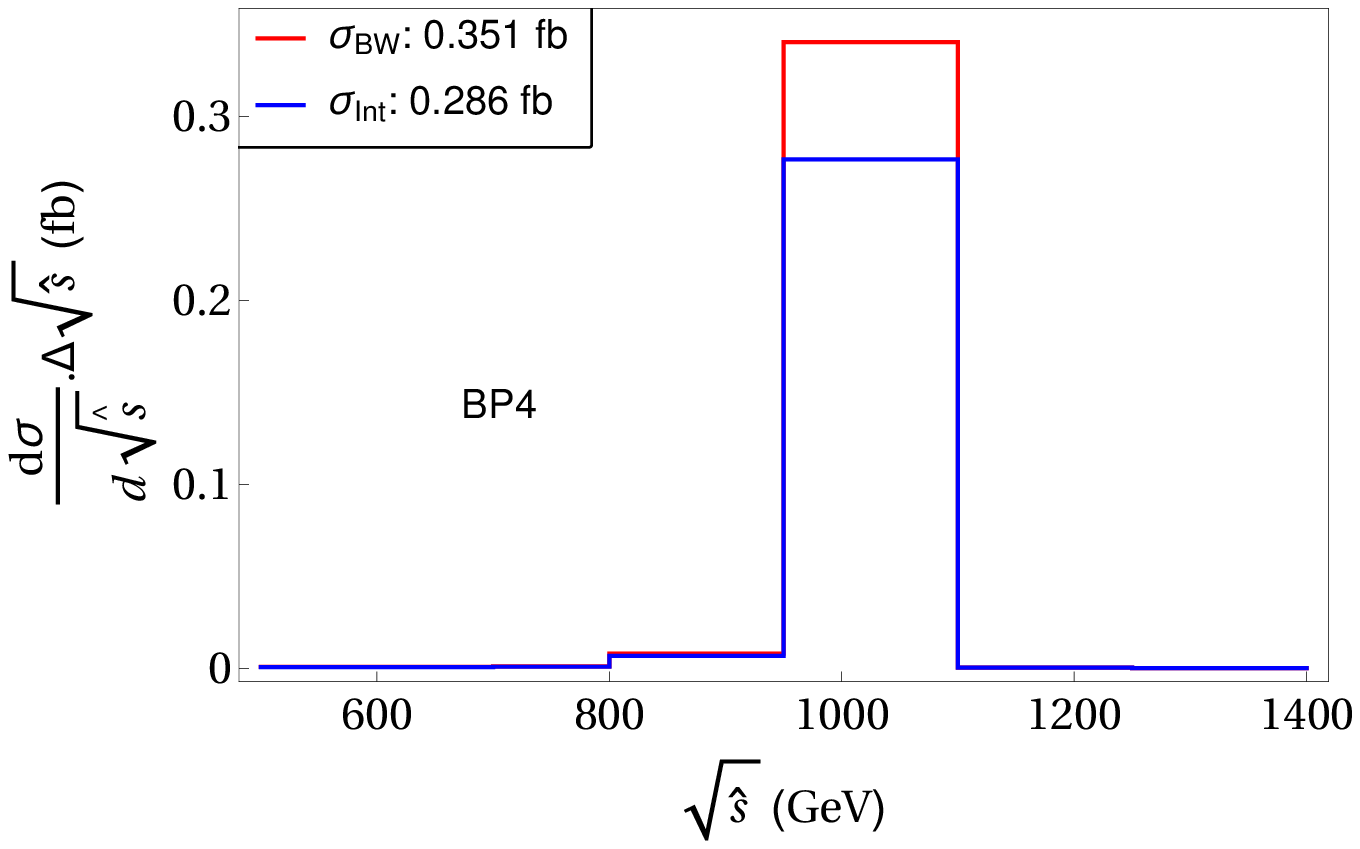} \\
\includegraphics*[width=7.8cm]{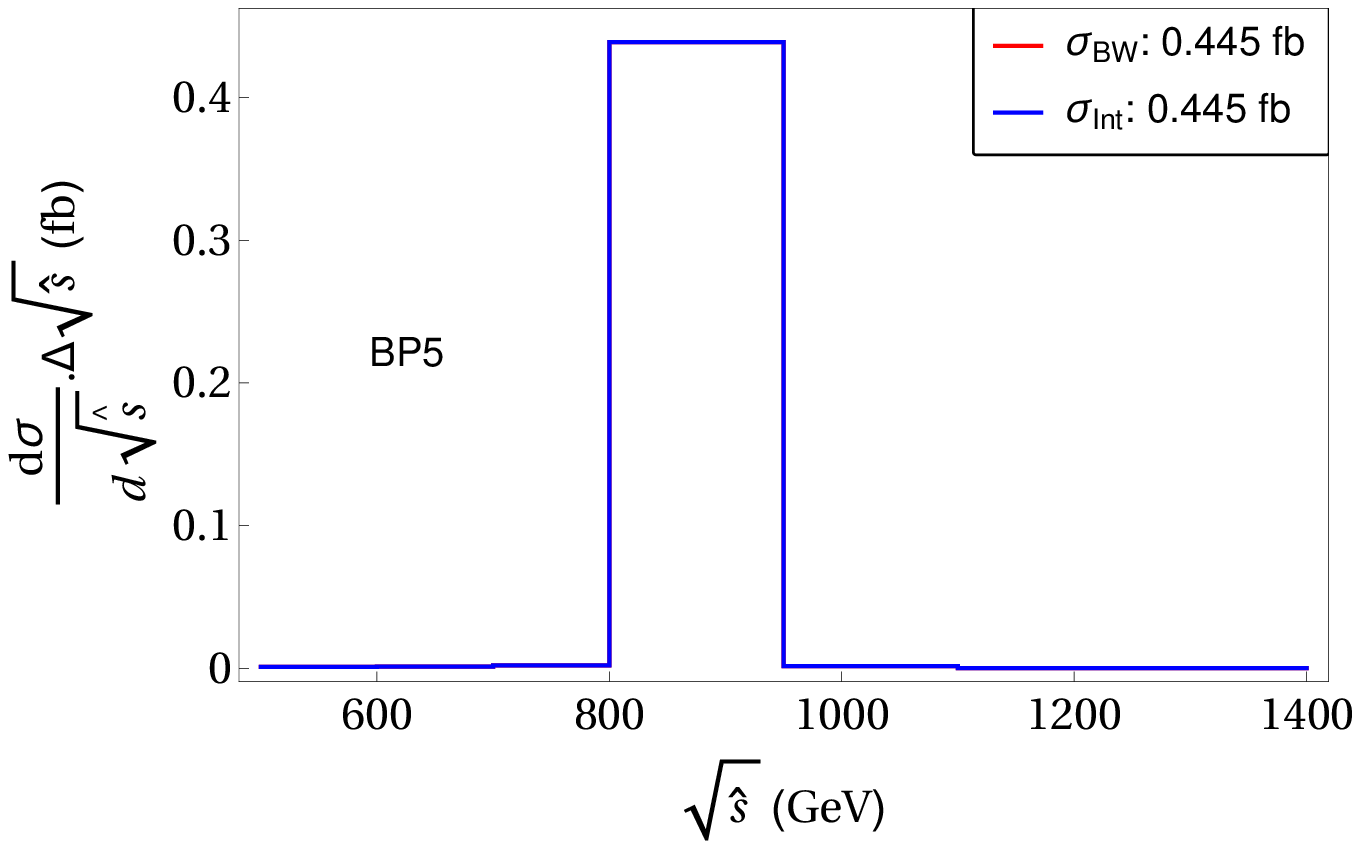} &
\includegraphics*[width=8cm]{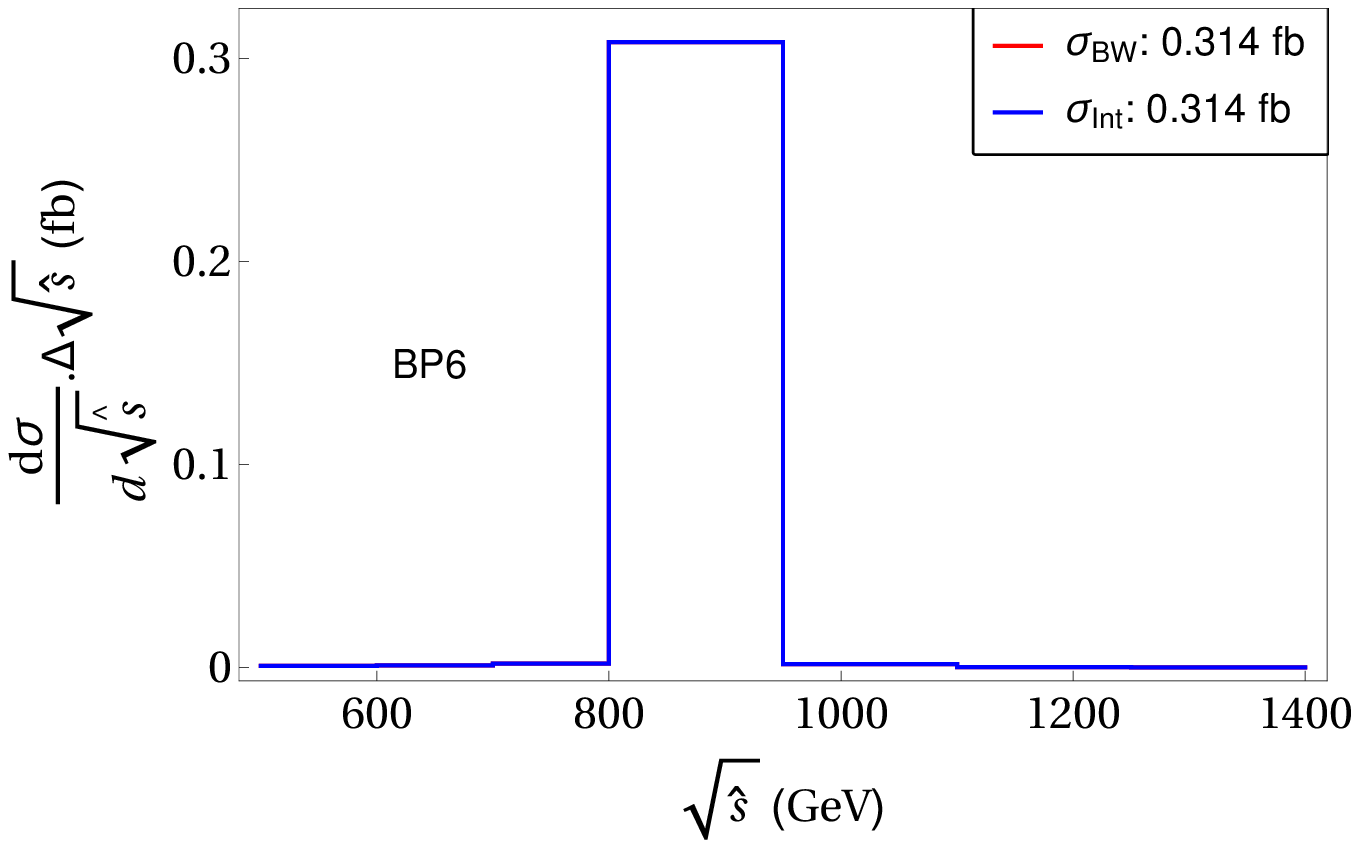} \\
\end{tabular}
\caption{\label{fig:bps-distri} Distributions of the 
differential cross sections with respect to $\sqrt{\hat s}$ 
for the six selected BPs. The blue lines correspond to the amplitude
containing the full Higgs propagator matrix, and the red lines to
the one assuming individual BW propagators.}
\end{figure}

Next, we also calculated the differential $\sigma_{\rm BW}$ and $\sigma_{\rm Int}$ with respect to the $\sqrt{\hat s}$, in order to examine if the interference effects can lead to  visible differences between the shapes of their distributions and may therefore be possible to probe at the LHC. For these we employ a binning template closely replicating the one used by the ATLAS collaboration in the searches for heavy resonances in the $\tau^+\tau^-$ decay channel~\cite{Aaboud:2017sjh}, based on an expected detector mass resolution of $\sim 15-20\%~{\rm of}~M_{\tau^+\tau^-}$. Our template thus assumes bins of width 50\,GeV, 100\,GeV and 150\,GeV for $\sqrt{\hat s} = 0-500\,{\rm GeV},~500-800\,{\rm GeV}~{\rm and}~800-1400\,{\rm GeV}$, respectively. The differential distributions are shown in figure~\ref{fig:bps-distri} for all the six BPs, with the red lines corresponding to $\sigma_{\rm BW}$ and the blue ones to $\sigma_{\rm Int}$. Since the mass-degenerate Higgs bosons in both the scenarios investigated are always heavier than 800\,GeV and the remaining two Higgs bosons typically much lighter, these distributions have a lower cut-off at $\sqrt {\hat s} = 500$\,GeV. In fact, for these distributions we included only $H_i$ with $i=3-5$ in the calculation of the two differential cross sections. 

Importantly though, the two differential distributions do not reveal much beyond what can be inferred from the total cross sections, owing to the poor $M_{\tau^+\tau^-}$ resolution at the LHC. We noticed earlier that $\Delta m_H$ in scenario-1 never exceeds 4\,GeV, and even this maximum value is about 1/38th of the adopted widths of the bins around $\sqrt {\hat s} = 1$\,TeV, despite $\sim 0.15 \times M_{\tau^+\tau^-}$ itself being on the conservative side of the expected detector resolution. Thus, the red boxes in each of the first four panels of figure~\ref{fig:bps-distri} rise higher than the blue ones (recall that the interference is always destructive) only in the bins within which the three mass-degenerate Higgs bosons lie. In the remaining bins, the boxes in the two histograms almost entirely overlap. Evidently, for BP5 and BP6 from scenario-2, a complete overlapping of the two lines occurs even in the bin containing the three heavy Higgs bosons (so that the red lines, plotted first, are entirely invisible behind the blue ones), since no interference effects appear there.   

\begin{figure}[t!]
\begin{tabular}{cc}
\includegraphics*[width=7.8cm]{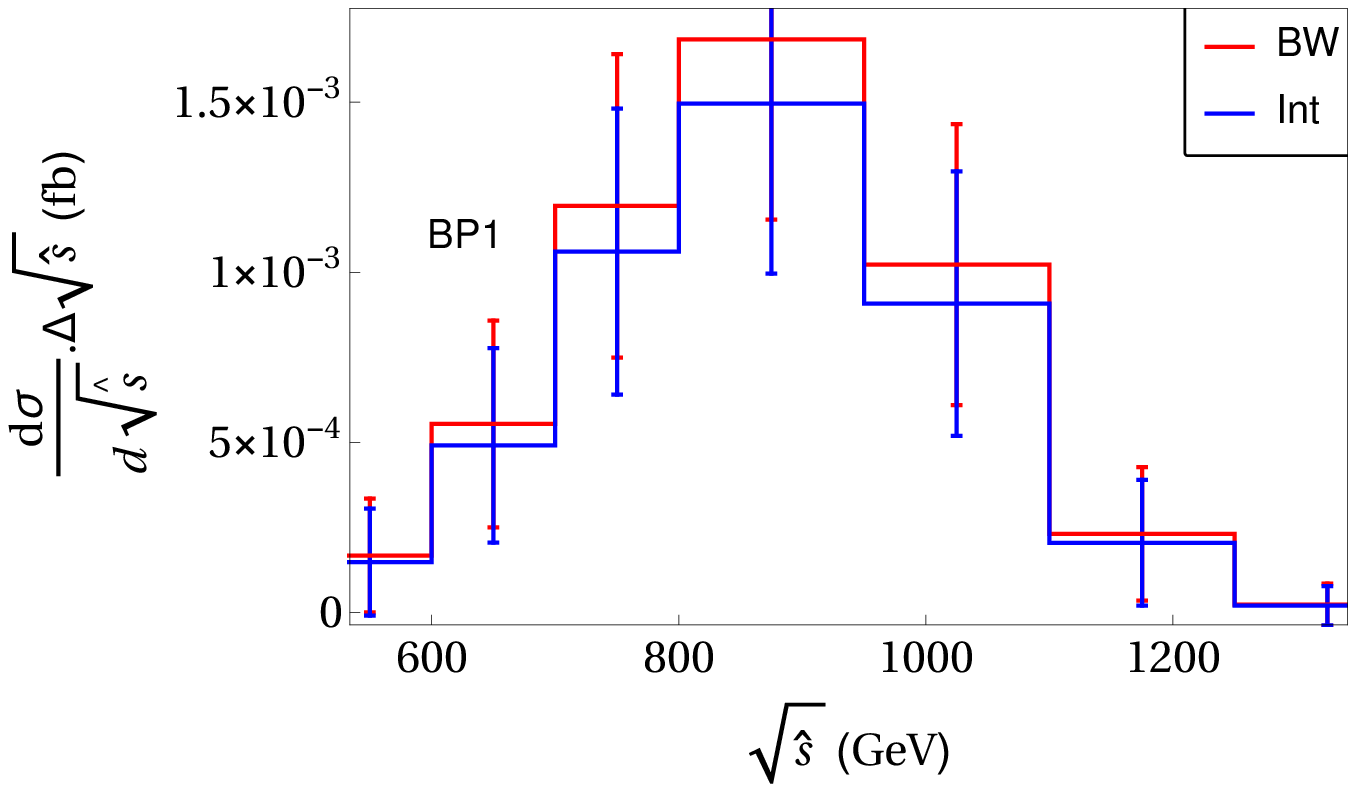} &
\includegraphics*[width=7.8cm]{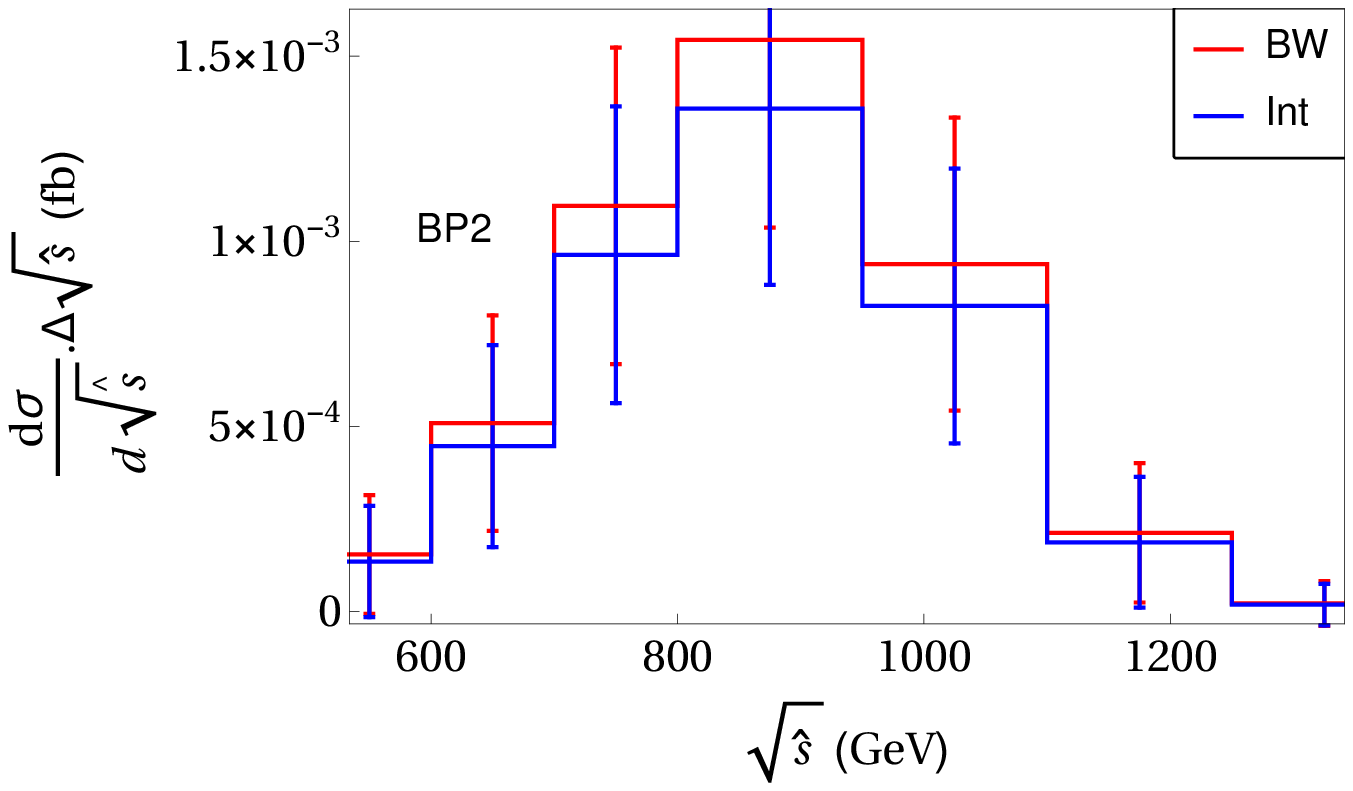} \\
\includegraphics*[width=7.8cm]{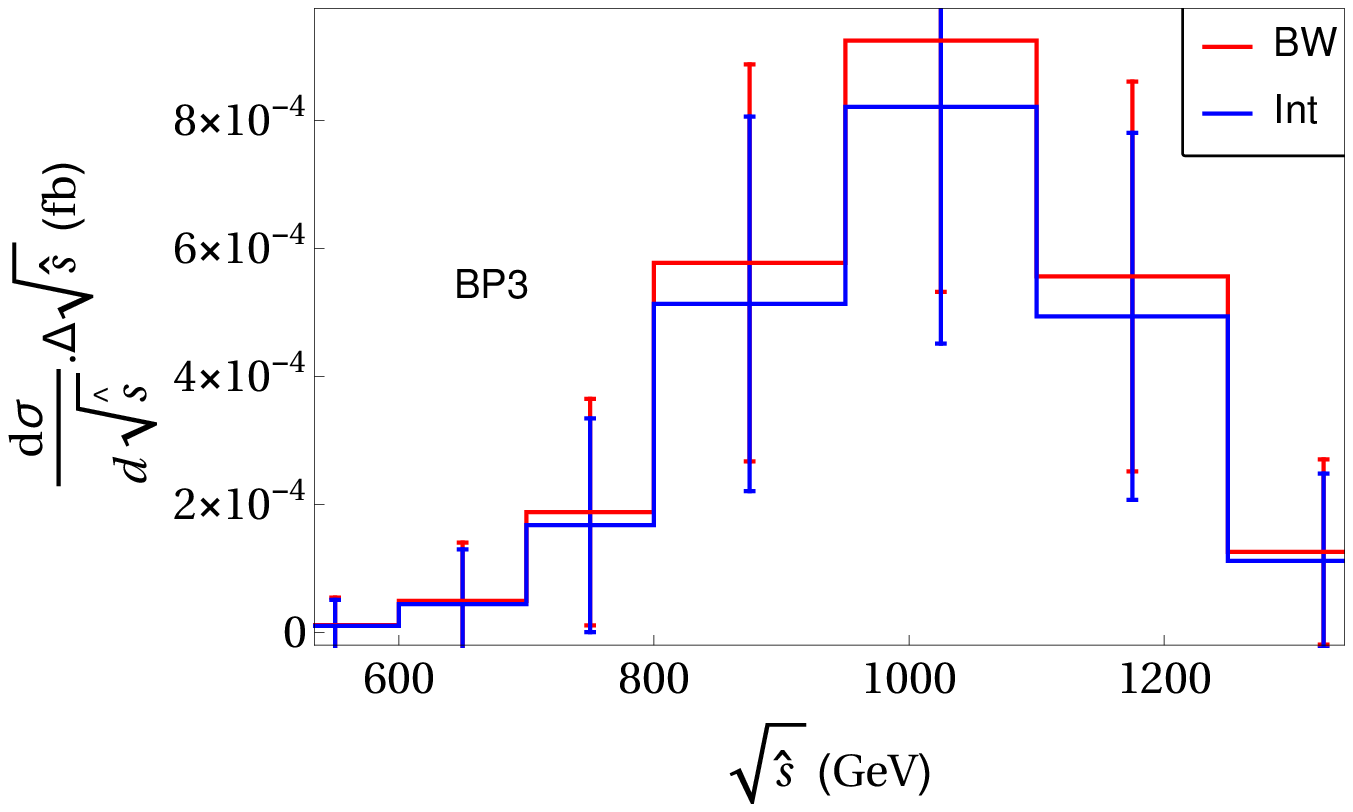} &
\includegraphics*[width=7.8cm]{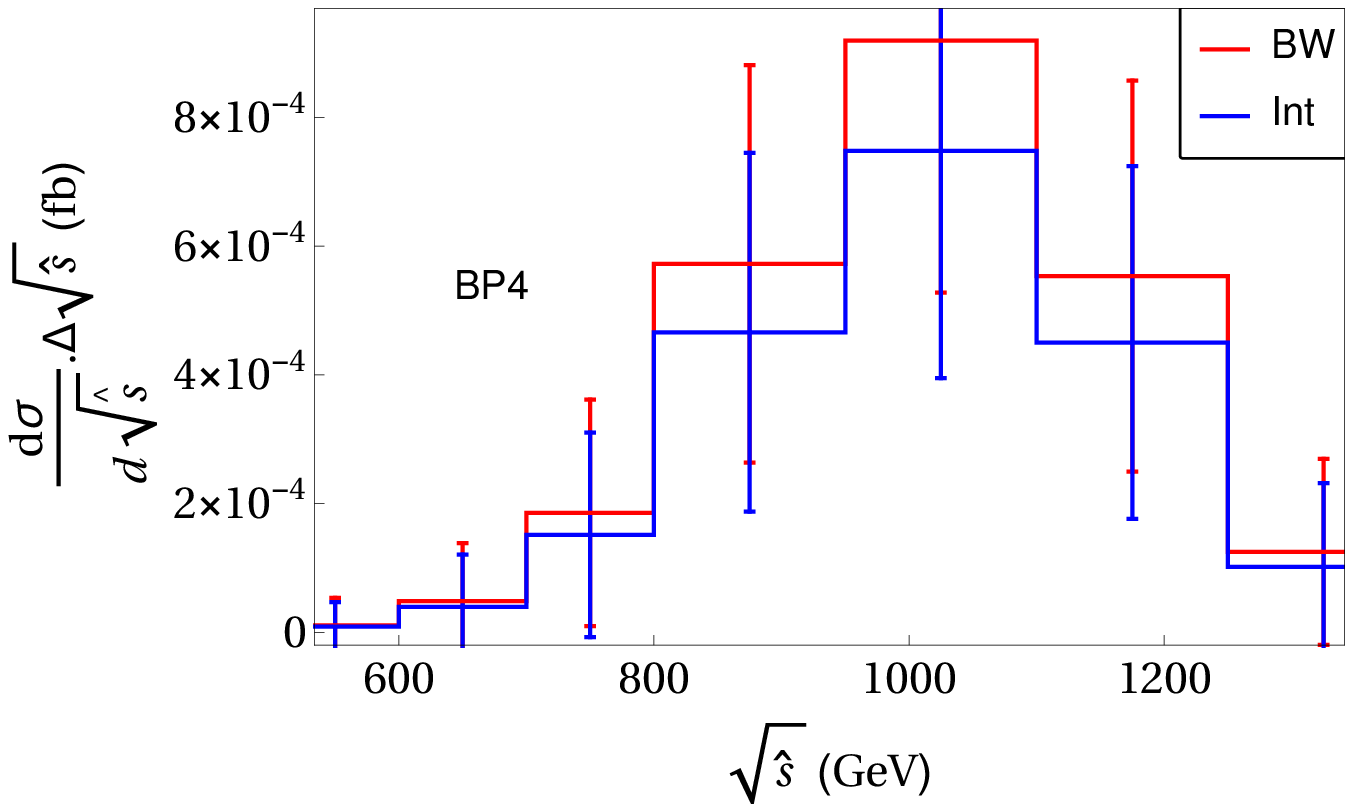} \\
\end{tabular}
\caption{\label{fig:bp-convo} Distributions of the 
differential cross sections for the four selected BPs of scenario-1,
after convolution with Gaussians of width 150\,GeV. The color
convention for the lines is the same as in figure\,\ref{fig:bps-distri}, and the error bars on them correspond to an
assumed integrated luminosity of 6000\,fb$^{-1}$.}
\end{figure}

Finally, we took the distributions for the differential $\sigma_{\rm BW}$ and $\sigma_{\rm Int}$ for the four BPs of scenario-1 and convolved them with a Gaussian of 150\,GeV width, using the {\tt ListConvolve} function~\cite{mathematica} in {\tt Mathematica}. The purpose of this exercise is to emulate the detector effect of smearing the cross sections over a few adjacent bins, to see whether the resulting distributions can be visually distinguishable from each other, despite the poor mass resolution. Figure~\ref{fig:bp-convo} shows these convolved distributions. We note that the convolution spreads out the differences in the heights of the red and blue boxes to the bins around the ones containing the resonant Higgs masses. However, the shapes of these convolved distributions for $\sigma_{\rm Int}$ do not show any peculiar features. They could very well be the distributions of the differential $\sigma_{\rm BW}$ for slightly different parameter space points, with the peaks arising from a single heavy Higgs resonance or even from nearly mass-degenerate $H$ and $A$ (with the underlying model being simply the MSSM). Therefore, even with an integrated luminosity as large as 6000\,fb$^{-1}$ (as assumed in this figure, in order to reduce the sizes of the error bars, which account for the statistical error only), the LHC will not be able to exploit the interference effects in order to identify two Higgs resonances with highly identical masses.

\section{\label{sec:concl}Conclusions}

The commonly adopted approach of calculating
the cross section for a given  $2\to 2$ process by factorising it into
the production and decay parts, assuming a
narrow width of the mediator, by construction
cannot account for the possible quantum interference among the
propagators of several mass-degenerate states. 
In this study,
we have considered the specific example of the NMSSM, wherein nearly
identical-mass pairs of CP-even or CP-odd Higgs bosons are viable over substantial
regions of the parameter space. These regions were found by 
numerical scanning of broad ranges of the model parameters, while
imposing the most important experimental constraints, including those
from the LHC pertaining to the Higgs, exotic and flavour sectors, as
well as those from the DM searches. 

By analysing six illustrative benchmark points from the
scanned set, we have highlighted
the importance of taking the interference effects into account. This was done
by including the full propagator matrix in
the calculation of the cross section for the process of production of
$\tau^+\tau^-$ in gluon fusion at the 14\,TeV LHC. We have shown that this cross
section can deviate considerably from the one obtained by employing the NWA, and even from the one obtained assuming BW propagators, an approach most
experimental searches are based on. This deviation, in fact, implies a
reduction in the cross section in the case of two mass-degenerate
CP-even Higgs bosons, as the interference is always destructive. In
the case of CP-odd states, on the other hand, no interference effects appear. We have also reasserted the fact that the smaller the
mass-splitting between two nearby Higgs bosons compared to the sum of
their widths, the larger the interference effects. 

The reason for considering the $\tau^+\tau^-$ decay channel of
the Higgs bosons is that the $\gamma\gamma$ channel, while cleaner,
has a prohibitively small decay rate for Higgs bosons with masses
close to 1\,TeV, The BR for the $\tau^+\tau^-$ is significantly
larger, but the price to pay is a poor experimental resolution
and mass reconstruction. As a result of which, we have concluded that the LHC
will be unable to disentangle the two resonances, even if they have a
mass splitting of a few GeV and the integrated luminosity is as large as
6000\,fb$^{-1}$. Thus, even though the $gg\to\tau^+\tau^-$ channel is an effective search instrument for heavy Higgs bosons, in the particular scenario when the latter are highly degenerate in mass, the former is unable be resolve them. Alternative Higgs boson production and decay modes that might be more suitable for this purpose are therefore the subject of a future study.

\section*{Acknowledgments}

SMo is supported in part through the NExT Institute, the STFC Consolidated Grant ST/L000296/1 and the H2020-MSCA-RISE-2014 grant no. 645722 (NonMinimalHiggs). We thank Korea Institute for Advanced Study for providing computing resources (Linux Cluster System at KIAS Center for Advanced Computation) for this work. 


\ifx\mcitethebibliography\mciteundefinedmacro
\PackageError{unsrtM.bst}{mciteplus.sty has not been loaded}
{This bibstyle requires the use of the mciteplus package.}\fi

\end{document}